\documentclass[12pt,preprint]{aastex}

\newcommand{\kms}{{\rm km~s}\ensuremath{^{-1}}}
\newcommand{\mrk}{Mrk~1383}
\newcommand{\pks}{PKS~2005--489}

\newcommand{\CII}{\ion{C}{2}}
\newcommand{\CIII}{\ion{C}{3}}
\newcommand{\CIV}{\ion{C}{4}}
\newcommand{\HI}{\ion{H}{1}}
\newcommand{\NV}{\ion{N}{5}}
\newcommand{\OVI}{\ion{O}{6}}
\newcommand{\SiII}{\ion{Si}{2}}
\newcommand{\SiIII}{\ion{Si}{3}}
\newcommand{\SiIV}{\ion{Si}{4}}

\newcommand{\NCII}{\ensuremath{N_{\rm C\,II}}}
\newcommand{\NCIII}{\ensuremath{N_{\rm C\,III}}}
\newcommand{\NCIV}{\ensuremath{N_{\rm C\,IV}}}
\newcommand{\NHI}{\ensuremath{N_{\rm H\,I}}}
\newcommand{\NNV}{\ensuremath{N_{\rm N\,V}}}
\newcommand{\NOVI}{\ensuremath{N_{\rm O\,VI}}}
\newcommand{\NSiII}{\ensuremath{N_{\rm Si\,II}}}

\newcommand{\NSiIV}{\ensuremath{N_{\rm Si\,IV}}}

\begin{document}

\title{Does the Milky Way Produce a Nuclear Galactic Wind?}
\author{Brian A. Keeney, Charles W. Danforth, John T. Stocke, Steven V. Penton, and J. Michael Shull}
\affil{Center for Astrophysics and Space Astronomy, Department of Astrophysical and Planetary Sciences, Box 389, University of Colorado, Boulder, CO 80309; keeney@casa.colorado.edu, danforth@casa.colorado.edu, stocke@casa.colorado.edu, spenton@casa.colorado.edu, mshull@casa.colorado.edu}
\author{\and Kenneth R. Sembach}
\affil{Space Telescope Science Institute, 3700 San Martin Drive, Baltimore, MD 21218; sembach@stsci.edu}

\shorttitle{Nuclear Galactic Wind}
\shortauthors{Keeney et al.}

\begin{abstract}

We detect high-velocity absorbing gas using {\em Hubble Space Telescope} and {\em Far Ultraviolet Spectroscopic Explorer} medium resolution spectroscopy along two high-latitude AGN sight lines (\mrk\ and \pks) above and below the Galactic Center (GC). These absorptions are most straightforwardly interpreted as a wind emanating from the GC which does {\em not escape} from the Galaxy's gravitational potential. Spectra of four comparison $B$ stars are used to identify and remove foreground velocity components from the absorption-line profiles of \OVI, \NV, \CII, \CIII, \CIV, \SiII, \SiIII, and \SiIV. Two high-velocity (HV) absorption components are detected along each AGN sight line, three redshifted and one blueshifted. Assuming that the four HV features trace a large-scale Galactic wind emanating from the GC, the blueshifted absorber is falling toward the GC at a velocity of $250\pm20$~\kms, which can be explained by "Galactic fountain" material that originated in a bound Galactic wind. The other three absorbers represent outflowing material; the largest derived outflow velocity is $+250\pm20$~\kms, which is only 45\% of the velocity necessary for the absorber to escape from its current position in the Galactic gravitational potential. All four HV absorbers are found to reach the same maximum height above the Galactic plane ($|z_{\rm max}| = 12\pm1$~kpc), implying that they were all ejected from the GC with the same initial velocity. The derived metallicity limits of $\gtrsim 10$--20\% Solar are lower than expected for material recently ejected from the GC unless these absorbers also contain significant amounts of hotter gas in unseen ionization stages.

\end{abstract}

\keywords{Galaxy: center --- intergalactic medium --- ISM: clouds --- ISM: jets and outflows --- quasars: absorption lines}

\section{Introduction}
\label{intro}

The presence of a powerful nuclear wind would have a significant impact on the Galactic environment. A Galactic wind could be responsible for the early enrichment of the halo and outer disk implied by the observed chemical abundances of the thick disk and globular clusters \citep{freeman02}. A powerful nuclear wind could also constrain the size of the Galactic bulge \citep{carlberg99} and influence the evolution of the Milky Way dwarf spheroidals \citep{irwin87}. 

More generally, galactic winds are a leading mechanism for transporting metals and energy from galaxies to the intergalactic medium (IGM), although it is unclear which galaxies are primarily responsible for IGM enrichment. Luminous, massive galaxies have higher rates of star formation and metal production than dwarf galaxies, so they have more material available for IGM enrichment. Typical luminous starbursts have star formation rates (SFRs) of $\sim 1$--10~$M_{\Sun}$~yr$^{-1}$ \citep*{heckman90}, although mergers can create ultraluminous galaxies with SFRs up to 100~$M_{\Sun}$~yr$^{-1}$ \citep{martin03}. Nearby dwarf starbursts have typical SFRs on the order of 0.1--1~$M_{\Sun}$~yr$^{-1}$ \citep{martin03}. However, \citet{meurer97} find a maximum areal SFR of $\sim 45$~$M_{\Sun}$~kpc$^{-2}$~yr$^{-1}$ for starbursts of all luminosities. This result implies that dwarf starbursts may be more efficient at transporting the metals entrained in their winds to the IGM because of their shallower gravitational potentials with lower escape velocities than massive starbursts with massive winds. If this latter view is correct, our own Galaxy's wind may not escape to enrich Local Group gas.

There have been several indications that our Galaxy has a nuclear wind. \citet{lockman84} found an absence of \HI\ 21~cm emission at $|z| > 500$~pc and $R \lesssim 3$~kpc from the Galactic center (GC), which he attributed to a clearing of the neutral gas by a nuclear wind. More recently, powerful mass ejections from the GC have been observed on scales of several arcminutes to tens of degrees in the infrared, radio, X-rays, and $\gamma$-rays \citep*{morris96,yusef-zadeh00,cheng97}. Evidence for a large-scale bipolar wind emanating from the GC was found in infrared dust emission by the {\em Midcourse Space Experiment} \citep[MSX;][]{bland-hawthorn03}. Observations in the MSX 8.3~\micron\ band discovered a limb-brightened bipolar structure at the GC that extends $\gtrsim 1\degr$ on either side of the plane, which \citet{bland-hawthorn03} attributed to dust entrained in a Galactic wind powered by a central starburst several million years ago. \citet{bland-hawthorn03} argue that the amount of energy required to entrain large amounts of molecular gas and dust into the region observed by MSX is of the same order as the amount of energy required to explain the position of the North Polar Spur (NPS) if it arose from a nuclear explosion. Thus, they suggest that the GC drives large-scale winds into the halo every $\sim 10$--15~Myr.

Starburst winds have been studied in nearby galaxies in H$\alpha$ and X-ray emission \citep*{watson84,martin99,martin02} and in UV and optical absorption \citep{heckman00, heckman01, keeney05}, but both techniques have limitations. Emission-line studies measure the extent and temperature of the outflowing wind only in its densest regions. They cannot measure the wind velocity in the diffuse halo nor the full extent of the wind gas to determine whether the wind material is bound to the galaxy. Detailed studies of the absorption-line features produced by starburst winds require a bright background source, typically the stellar continuum of the starburst region itself. Absorption-line studies are much more sensitive to diffuse gas than emission-lines studies, but they only yield information about one line of sight, and they suffer from an ambiguity in the distance between the background source and the absorbing gas.

Wind studies that use the stellar continuum of the starburst as a background source cannot distinguish a high-velocity outflow in the galactic halo from one in the starburst region itself. This ambiguity complicates the interpretation of the typical outflow velocities found by these studies (400--1000~\kms), although they can be much smaller for dwarf galaxies \citep{heckman00, martin03}. It is not straightforward to use outflow velocities measured near the galactic disk to determine if the wind is gravitationally bound to the galaxy because the wind will decelerate under the influence of both gravity and mass-loading. This complication can be alleviated by measuring the outflow velocity well away from the host galaxy. Currently, this can only be done if there is a bright background QSO close to the galaxy of interest \citep[e.g.,][]{stocke04, keeney05,keeney06}. 

Determining whether a starburst wind is gravitationally bound to its host galaxy is essential for assessing which galaxies are responsible for the metal enrichment of the IGM. \citet{martin99} found that the temperature of starburst winds is nearly constant as a function of a galaxy's maximum \HI\ rotation speed (i.e., mass), implying that the strength of a starburst wind is independent of galaxy luminosity. \citet{stocke04} found that the unbound wind produced by a dwarf poststarburst galaxy located 71$\;h_{70}^{-1}$~kpc from the sight line could be responsible for the 1586~\kms\ metal-line system observed in the spectrum of 3C~273. On the other hand, \citet{keeney05} studied the nearby luminous ($0.5\,L^*$) starburst galaxy NGC~3067 (${\rm SFR} \approx 1.4~M_{\Sun}$~yr$^{-1}$) and found that its wind is bound along the line of sight to 3C~232, which probes the halo of NGC~3067 near its minor axis $11\;h_{70}^{-1}$~kpc from the plane. Thus, the results of \citet{stocke04} and \citet{keeney05} support the expectations of \citet{meurer97} and \citet{martin99} that winds can escape more readily from dwarf galaxies than from massive starbursts.

In this paper, we use UV absorption line spectroscopy obtained with the {\it Hubble Space Telescope} (HST) and the {\it Far Ultraviolet Spectroscopic Explorer} (FUSE) to detect high-velocity absorption that we interpret as originating in our own Galaxy's nuclear wind. We present spectra of two high-latitude AGN that probe either side of the GC where a nuclear wind would be expected: \mrk\ probes the northern axis at $(l,b) = (349\degr,55\degr)$ and \pks\ probes the southern axis at $(l,b) = (350\degr,-33\degr)$. We identify foreground absorption components from the nearby interstellar medium (ISM) and Sco-Cen OB Associations with four comparison $B$ stars, two near each AGN sight line. In \S\,\ref{obs} we describe our observations and data reduction process. Several models of the outflow geometry and the kinematics of the high-velocity absorbers detected toward the GC are discussed in \S\,\ref{hvkin}. In \S\,\ref{ionization} we compare the ionization states of these high velocity absorbers with those found in highly-ionized high velocity clouds. Our results are summarized in \S\,\ref{conclusion}.

\section{Observations and Data Reduction}
\label{obs}

Our dataset consists of FUSE and HST/Space Telescope Imaging Spectrograph (STIS) ultraviolet spectra of six lines of sight. The AGN sight lines are the best available to probe the regions of the halo above and below the Galactic Center (GC) due to their high Galactic latitudes, proximity to $l = 0\degr$, and suitably bright FUV fluxes. The comparison stars are located in the same regions of sky and were chosen for their locations, brightnesses, and reasonable FUV continua. Table~\ref{tab:targets} lists basic physical data for our two AGN and four comparison star sight lines. The velocity corrections required to convert the observed heliocentric wavelength scale to the local standard of rest (LSR), $\Delta v_{\rm lsr} \equiv v_{\rm lsr} - v_{\rm hel}$, are shown in Table~\ref{tab:targets} and assume a solar motion relative to the LSR of 16.5~\kms\ toward $(l,b) = (53\arcdeg,25\arcdeg)$ \citep{mihalas81}. Figure~\ref{fig:posn} shows the relative locations of the AGN and comparison star sight lines and Figure~\ref{fig:geometry} displays a schematic diagram of the  sight line positions in Galactic coordinates. The AGN ESO~141--G55 is also shown in Figure~\ref{fig:posn} because it lies close to \pks\ and HD~191466 on the sky and was found by \citet{sembach03} to have high velocity \OVI\ absorption \citep*[and may contain some high velocity \NV\ and \CIV\ absorption as well;][]{penton00,indebetouw04b} at similar velocities to those seen in \pks. However, because ESO~141--G55 is further from the GC than \pks, we rely on the published studies for information on its HV absorbers."

A journal of the FUSE and HST/STIS observations appears in Table~\ref{tab:obslog}, which lists the instrument, grating, dataset ID, wavelength coverage, spectral resolution, and the total usable exposure time for each target. Besides obtaining new HST/STIS and FUSE observations for this program, we have utilized all available data on these targets from the FUSE and HST/STIS archives. Data were obtained in all four FUSE channels \citep{moos00,sahnow00} for all of the targets in Table~\ref{tab:obslog} except for HD~191466, which has only side 1 data. Originally, HST/STIS E140M exposures were approved and planned for all targets in Table~\ref{tab:targets}. However, the STIS hardware failure in the summer of 2004 prevented the completion of this project.

FUSE data were retrieved from the archive and reduced locally using {\sc calfuse}v2.4\footnote{More detailed information is available at {\tt http://fuse.pha.jhu.edu/analysis/calfuse\_intro.html}.}. Raw exposures within a single FUSE observation were coadded by channel mid-way through the pipeline. This can produce a significant improvement in data quality for faint sources (such as AGN) since the combined pixel file has higher signal-to-noise than the individual exposures and consequently the extraction apertures are more likely to be placed correctly. Combining exposures also speeds up reduction time dramatically.  Reduced data were then shifted and coadded by observation to generate a final spectrum from each of the eight data channels. The data were binned by three pixels; FUSE resolution is typically 8--10 pixels or roughly 3 bins. The reduced FUSE data were normalized in 10~\AA\ segments around \CIII\ $\lambda977$, \OVI\ $\lambda1032$, and \OVI\ $\lambda1038$. Line-free regions of the stellar and AGN continua were selected interactively and fitted using Legendre polynomials of order less than six. The \pks\ data show moderate airglow emission blended with the \CIII\ $\lambda977$ line. We performed a special reduction using only data taken during orbital night (22.8~ksec) for the analysis of \CIII\ toward \pks\ but use the full 48.7~ksec for \OVI\ analysis.

STIS data were also retrieved from the archive and reduced locally. In addition to standard spectral processing, all STIS G140M and E140M data (see Table~\ref{tab:obslog}) were processed with a spectral deconvolution routine analogous to the ``CLEAN'' algorithm used in radio astronomy \citep{penton06}. This routine recovers the flux scattered in the broad wings of the line spread function (LSF) of the STIS $0\farcs2$ slit \citep*[see Fig.~1 of][]{penton04} by iteratively removing the deepest absorption feature in the spectrum and replacing it with a CLEANed component. The location of the deepest absorption feature in the input spectrum is found and the known LSF of the input grating+aperture is used to remove the feature's flux, including the contributions from the non-Gaussian wings, from the input spectrum. The component removed from the input spectrum is then inserted at the same location in the output spectrum, but with a Gaussian LSF that has the same FWHM as the LSF of the smallest aperture that can be used with the input grating (i.e., the $0\farcs05$ slit for the STIS first order gratings and the $0\farcs03$ slit for the echelle gratings). This process is repeated until a stop condition is met, at which point the residuals from the input spectrum are added to the output spectrum and the routine exits. This algorithm conserves the equivalent width of all CLEANed components and produces an output spectrum with deeper absorption features than the input spectrum due to a well-behaved Gaussian LSF instead of one in which a significant percentage of the flux is scattered into broad wings. However, the additional processing also introduces noise to the continuum. An input spectrum with a signal-to-noise ratio of 5 per pixel will have continuum fluctuations in the CLEANed spectrum that are 2\% larger than in the input spectrum, and the amount of noise added by the routine decreases with increasing signal-to-noise of the input spectrum \citep{penton06}. Inspection of these spectra before and after CLEAN processing reveals no great differences; i.e., the use of this processing does not alter the basic results presented here. After processing the G140M and E140M data with the CLEAN algorithm, the continua within 2000~\kms\ of \SiIII\ $\lambda1207$, \NV\ $\lambda\lambda1239, 1243$, \SiII\ $\lambda1260$, \SiIV\ $\lambda\lambda1394, 1403$, \CII\ $\lambda1335$, and \CIV\ $\lambda\lambda1548, 1551$ were normalized using Legendre polynomials fitted to line-free regions of the stellar or AGN continuum for all STIS spectra.

All spectra have been shifted to the LSR frame in the direction of the AGN by identifying strong low-ionization Galactic absorption lines that are present in the spectra of the AGN and each of their comparison stars. The spectra were then shifted until the chosen absorption line --- \ion{O}{1} $\lambda976$ in the \CIII\ region, \ion{O}{1} $\lambda1039$ or H$_2$ $\lambda\lambda1038.2, 1038.7$ in the \OVI\ region, and \ion{S}{2} $\lambda1251$ in the STIS data --- was centered at $v_{\rm lsr} = -2$~\kms\ for \mrk, HD~121968, and HD~125924 \citep[the LSR velocity in the direction of \mrk;][]{wakker03} and $v_{\rm lsr} = +2$~\kms\ for \pks, HD~187311, and HD~191466 \citep[the LSR velocity towards \pks;][]{wakker03}.  This method assures that the AGN and their comparison stars have a common velocity scale and allows us to ascertain which absorption components are present only in the AGN spectra. Since the comparison star spectra are only used for this purpose, the small discrepancy between the LSR velocity in the direction of the comparison stars as compared to the AGN are ignored.

Overplotting the absorption line profiles of \CII\ $\lambda1036$ and \CII\ $\lambda1335$ in the four sight lines for which we have data longward of 1300~\AA\ (see Table~\ref{tab:obslog}) indicates that the FUSE and STIS velocity scales are aligned to better than 6~\kms. Comparing the LSR velocities found by this method with the predictions of \citet[see Table~\ref{tab:targets}]{mihalas81} yields maximum differences of $33\pm2$~\kms\ in the FUSE data, $12\pm3$~\kms\ in the STIS/G140M data, and $8\pm2$~\kms\ in the STIS/E140M data. Systematic velocity shifts caused by centering errors in a wide aperture/slit are primarily responsible for this discrepancy; the FUSE LWRS aperture has a width of 100~\kms\ and the STIS $0\farcs2$ slit has a width of 80(20)~\kms\ when used with the G140M(E140M) grating.

The absorption line profiles of \OVI, \NV, \CII, \CIII, \CIV, \SiII, \SiIII, and \SiIV\ were fitted with multi-component Voigt profiles using a $\chi^2$ minimization routine. All spectra were smoothed to one resolution element per pixel before fitting, and the fits were allowed to vary such that the column density and Doppler $b$ value fell in the range $N = 10^{10-16}$~cm$^{-2}$ and $b = 2$--100~\kms. Hereafter, all column densities, $N$, will be quoted in units of cm$^{-2}$. Rest wavelengths, oscillator strengths, and transition rates were taken from \citet{morton03}. For heavily saturated profiles such as the \CIII\ profile in the \pks\ sight line (see Figure~\ref{fig:datafit}), the minimum number of components satisfying the above constraints were fit to the saturated regions. All components fit to \NV\ $\lambda1239$, \SiIV\ $\lambda1394$, and \CIV\ $\lambda1548$ were confirmed in the profiles of \NV\ $\lambda1243$, \SiIV\ $\lambda1403$, and \CIV\ $\lambda1551$. Unfortunately, components fit to \OVI\ $\lambda1032$ could not be confirmed in the same manner because the region near \OVI\ $\lambda1038$ is contaminated with strong \CII\ $\lambda1036$ and \CII* $\lambda1037$ absorption, as well as several H$_2$ lines. The \CIII\ $\lambda977$ transition has a neighboring \ion{O}{1} transition at 976.4~\AA. To eliminate ambiguity as to which velocity components are associated with the \ion{O}{1} line and which with \CIII, we assumed the \ion{O}{1} profile to have the same velocity components found in the \ion{O}{1} $\lambda1039$ line.

The only lines fitted in the HD~125924 sight line were \CIII\ and \OVI\ because no STIS spectrum is available for this target. Similarly, \SiIV\ and \CIV\ were not fitted in the spectra of \pks, HD~187311, or HD~191466 because the STIS G140M spectrum of \pks\ does not extend redward of 1300~\AA\ (see Table~\ref{tab:obslog}). \CII\ $\lambda1036$ was not fitted in the comparison stars owing to difficulty identifying the stellar continuum and blending of \CII\ $\lambda1036$ with nearby \CII* $\lambda1037$ and \OVI\ $\lambda1038$. The profile of \CII\ $\lambda1335$ was fitted in the E140H spectrum of HD~121968. However, while broad, saturated photospheric absorption was found centered at $v_{\rm lsr} \approx 0$ (extending from $-90$ to $+50$~\kms), a narrow component centered at $-60$~\kms\ was also detected in this spectrum. \SiIII\ was not fitted in the spectra of HD~187311 and HD~191466 because stellar \SiIII\ and Ly$\alpha$ absorption made continuum placement extremely uncertain. Only the high signal-to-noise ratio and spectral resolution of the HD~121968 data allowed us to identify line-free continuum regions and discriminate between interstellar and photospheric \SiIII\ absorption. \NV\ was not fitted in the HD~187311 and HD~191466 sight lines because all absorption features in the vicinity of \NV\ $\lambda\lambda 1239,1243$ in these spectra are photospheric. The \NV\ $\lambda1239$ feature in the HD~121968 sight line is very weak, but it is confirmed by a feature of the appropriate strength in \NV\ $\lambda 1243$.

Figure~\ref{fig:datafit} shows the absorption lines in the AGN spectra with the best-fit Voigt profiles overlaid as solid lines. The best-fit profiles for the comparison star spectra are shown as dashed and dotted lines, although the data themselves are not shown for these sight lines.

Since all of the comparison stars are within 4~kpc of the Sun (see Table~\ref{tab:targets}), all velocity components detected in these sight lines are associated with nearby interstellar material. Thus, any common absorption components in the AGN and its comparison stars represent foreground material that is not part of a Galactic wind. We have identified those AGN velocity components that have no overlap with any comparison star components as high-velocity (HV) components. We have classified the \NV\ component at $v_{\rm lsr} = -103\pm5$~\kms\ in the \pks\ sight line as a HV component, even though the \NV\ region is contaminated by photospheric absorption in the comparison stars for this sight line, because there are two HV \OVI\ components toward \pks\ that bracket it in velocity. However, we emphasize that, while the velocity correspondence is suggestive, we cannot definitively classify this as a HV component without knowing the extent of foreground \NV\ absorption. The HV components are indicated by the shaded absorption profiles in Figure~\ref{fig:datafit}, and the dotted vertical lines show the average velocity of these components over all ions in which they are detected.

The best-fit LSR velocities, $b$ values, column densities, and equivalent widths from our Voigt profile fits are shown in Table~\ref{tab:mrkparams} and Table~\ref{tab:pksparams} for HV absorption components in the \mrk\ and \pks\ sight lines, respectively. The velocity errors shown in these Tables reflect the errors in the line centroid from the fitting routine and the uncertainty in the LSR velocity calibration, combined in quadrature. All other errors are $1\sigma$ fitting errors. Tables~\ref{tab:mrkparams} and \ref{tab:pksparams} show significant overlap in the HV component velocities from ion to ion, indicating that these ions probably coexist spatially along the AGN lines of sight. Two HV absorbers are detected along the \mrk\ sight line that probes the northern Galactic axis: absorber N1 is detected in \CIII, \CIV, \SiII, and \SiIV\ at an average LSR velocity of $+45\pm7$~\kms, and absorber N2 is detected in \CIII, \CIV, \SiII, \SiIV, and \OVI\ at an average LSR velocity of $+94\pm12$~\kms. We also detect two HV absorbers in the \pks\ sight line that probes the southern axis, absorber S1 in \NV\ and \OVI\ at a velocity of $-105\pm12$~\kms, and absorber S2 in \CIII, \SiII, and \OVI\ at a velocity of $+168\pm10$~\kms.

\section{HV Absorber Kinematics}
\label{hvkin}

At $|v_{\rm lsr}| \geq 90$~\kms, absorbers N2, S1, and S2 satisfy the definition of high velocity clouds (HVCs), while absorber N1 is an intermediate velocity cloud \citep[IVC;][]{wakker97}. Since the HV absorbers are detected in species such as \CIV, \SiIV, \NV, and \OVI\ and not in \HI\ 21~cm emission, they are also part of the subset of HVCs that are classified as highly-ionized \citep*{sembach99,collins04,collins05}. Owing to the detection of the HV absorbers in these highly ionized species and the proximity of our high-latitude AGN sight lines to $l=0\degr$, we hypothesize that the HV absorbers were created by an outflowing Galactic wind. In contrast to merely labelling the HV absorbers HVCs, a purely kinematic classification, this hypothesis allows us to test one possibility for the origins and dynamics of these absorbers.

The distances to the HV absorbers are reasonably well constrained and imply that they lie near the GC. Since we exclude all absorption components common to the AGN and their respective comparison stars, the HV absorbing gas must be more distant than the low-velocity ISM in the Galactic disk. Thus, the HV absorbers in the \mrk\ sight line must be farther than 3.6~kpc from the Sun ($|z| > 3.0$~kpc) and the \pks\ HV absorbers must be more than 2.4~kpc away ($|z| > 1.3$~kpc; see Table~\ref{tab:targets}). This places these absorbers well beyond any possible outflow associated with the Sco-Cen OB Associations \citep[$d \approx 110$--150~pc;][]{preibisch99}. 

While the data do not {\em strictly require} that the HV absorbers are associated with gas more distant than the comparison stars, the detection of four small clumps of HV gas in the AGN sight lines but not toward the comparison stars is highly unlikely. Since the AGN and their comparison stars are separated by several degrees on the sky (see Figure~\ref{fig:posn}), the HV absorbers could be caused by very nearby absorbers with small transverse sizes. Given the angular separations of the AGN and their comparison stars (Table~\ref{tab:targets}), the maximum transverse size for the HV absorbers at the distance of the comparison stars is 1.4~kpc for the \mrk\ absorbers (N1 and N2) and 0.8~kpc for the \pks\ absorbers (S1 and S2). If the HV absorbers were closer than the comparison stars and had larger transverse sizes then they would have been detected in the spectrum of at least one of the comparison stars. 

The sight line to the AGN ESO~141--G55 lies $14\degr$ from \pks\ and $7\degr$ from the comparison star HD~191466 on the sky at $(l,b) = (338\degr, -27\degr)$ (see Figure~\ref{fig:posn}). \citet{sembach03} measured \OVI\ absorption at $v_{\rm lsr} = +176$~\kms, offset from the centroid of absorber S2 by only $\sim 10$~\kms, with a column density of $\log{\NOVI} = 13.45\pm0.18$ \citep{sembach03}. \citet{penton00} found possible \NV\ absorption at $v_{\rm lsr} \sim -65$~\kms\ in HST/GHRS spectra of ESO~141--G55, and additional FUSE and HST/GHRS data show \OVI\ and \CIV\ absorption at the same velocity \citep[see Figure~1 of \citealp{wakker03} and Figure~1$a$ of][]{indebetouw04b}. The \pks\ sight line shows \NV\ and \OVI\ absorption at similar velocities, with \NV\ absorption components at $v_{\rm lsr} = -36\pm5$~\kms\ and $v_{\rm lsr} = -88\pm6$~\kms, and a \OVI\ component at $v_{\rm lsr} = -70\pm16$~\kms\ (see Table~\ref{tab:pksparams} and Figure~\ref{fig:datafit}). The detection of \OVI\ at identical redshifted velocities in these two extragalactic sight lines strongly suggests that absorber S2 is not a very small local cloud. While similarly bright targets are not available near \mrk, it seems quite contrived that these four HV absorbers would cover just the AGN sight lines and not the comparison stars unless the absorbing gas is situated beyond the comparison stars.

\citet{savage03} found that the distribution of \OVI\ in the Galactic halo can be roughly described by a patchy plane-parallel absorbing layer with a midplane column density of $\log{(\NOVI\,\sin{|b|})} = 14.09\pm0.25$, a scale height of $\sim 2.3$~kpc, and a $\sim 0.25$ dex excess of $\log{\NOVI}$ at $b > 45\degr$. The perpendicular column densities of our HV \OVI\ absorbers\footnote{The column densities of the HV O\,VI absorbers are also in good agreement with the value of $\langle\log{\NOVI}\rangle = 13.95\pm0.34$ found by \citet{sembach03} for a sample of 84 high-velocity O\,VI absorbers.} imply that they are 2--3 scale heights above the midplane ($|z| = 5$--7~kpc). This constraint places rough upper limits on the distances to the HV absorbers of $\sim 6$--9~kpc in the \mrk\ sight line and $\sim 9$--13~kpc in the \pks\ sight line, indicating that the HV absorbing gas is located near the GC or between the comparison stars and the GC. More stringent constraints on the distance to the HV absorbers are derived in \S\,\ref{kinmod} in the context of a specific geometrical model.

The above arguments place these HV absorbers between 3 and 10~kpc from the Sun. Why do we ascribe them specifically to the GC? First, there is the considerable indirect evidence for a GC wind described in \S\,\ref{intro}. Then there is the analogy with other galaxies, where winds are observed above and below star forming regions in galaxy disks, in general, and above and below galactic centers, in particular \citep{heckman00,heckman01,martin02}. Diverse evidence for a nuclear starburst near our own GC includes the discovery of three rich clusters of early-type stars which could account for nearly 10\% of all massive stars in our galaxy \citep{figer03}. One of these, the Arches Cluster, is thought to contain $10^4$~M$_{\Sun}$ of stars created in a burst some $2.5\pm0.5$~Myr ago and producing $> 10^{51}$ ionizing photons s$^{-1}$ \citep{figer02}. There is also evidence that these very young star clusters are only the latest in a series of starbursts at the GC \citep{figer02}. Beyond the Sco-Cen OB Associations, there is little evidence for active star forming sites between the Sun and the GC region, and while we cannot easily observe regions on the other side of the GC, it is unlikely that star formation in those regions is as intense as in the GC itself. An outflowing wind also could have been generated by the supermassive black hole at the GC \citep{proga03}, whose existence has been solidified by recent proper motion and Doppler shift observations of stars moving near it \citep{ghez03}. While the $(4.1\pm0.6) \times 10^6$~M$_{\Sun}$ black hole in our GC is small by Seyfert standards, it could neverless be responsible for an outflowing wind \citep[][and references therein]{melia01}. Given these strong indicators, it would have been surprising if we did {\em not} detect HV absorption in these directions.

Secondly, we have detected two HV absorbers along each sight line; none of these absorbers are associated with \HI\ 21~cm detected HVCs. Indeed, \citet{lockman84} found an absence of HVCs toward the GC, which he attributed to a clearing of neutral gas by a GC wind. \citet{sembach03} report that high-velocity \OVI\ absorptions cover $\sim 60$\% of the sky to absorber strengths comparable to those detected here. But most are associated with \HI\ column densities ($\NHI \gtrsim 10^{18.5}~{\rm cm}^{-2}$) much larger than in these clouds (e.g., known \HI\ 21~cm HVCs and spatial extensions of known HVCs at similar radial velocity ranges to the nearby HVCs) and thus potentially represent collisionally-ionized \OVI\ due to HVC/halo gas interactions \citep{fox05}. Thus, the probability that these two sight lines would both contain two \OVI\ absorbers and no \HI\ 21~cm emission is small unless there is a direct relationship to the GC, as suggested by \citet{lockman84}. The probability of finding two extragalactic \OVI\ absorbers (e.g., Local Group gas) along both sight lines seems equally remote, although it cannot be ruled out.

In conclusion, while we cannot strictly rule out locations for this absorbing gas between 3~kpc distance and the GC and well beyond the GC, these are unlikely. In subsequent sections we will present further evidence that bears on this point: dynamical evidence that favors a causal relationship with components of the GC (star-forming regions and/or black hole) and metallicity evidence that argues somewhat against this conclusion.

\subsection{Outflow Geometry}
\label{outflow}

If the HV absorbers originated in a nuclear wind, then the interpretation of their observed velocities depends on the wind geometry. Nearby starburst galaxies exhibit conic or cylindrical outflows \citep{bland88, gotz90, mckeith95, shopbell98} with typical cone opening angles of $\sim 65\degr$ \citep{heckman90}. Similar outflow geometries have also been predicted by numerical models \citep[e.g.,][]{tomisaka88,strickland00}. \citet{wang02} detected a biconic region of 1.5~keV X-ray emission extending to $b = \pm16\degr$ ($|z| \approx 2$~kpc, assuming $R_{\Sun} = 8.5$~kpc) in a ROSAT image of the GC. In order to explain this X-ray emission and the observed shape of the North Polar Spur (NPS), a thermal X-ray/radio loop that extends from $b=0$--$80\degr$, \citet{bland-hawthorn03} proposed a geometry in which a central conic outflow evolves into a cylindrical wind at large heights above the plane. In order for high-latitude objects such as the NPS to be projected to their observed shapes, the central conic outflow must transition to a cylindrical wind interior to the Solar circle \citep{bland-hawthorn03}.

Before our observed velocities can be interpreted in the context of a Galactic outflow, they must be converted from the LSR frame to the Galactic standard of rest (GSR) by removing the rotation of the LSR about the GC: $v_{\rm gsr} \equiv v_{\rm lsr} + (220~\kms) \cos{b} \sin{l}$. Since $l \approx 350\degr$ for both AGN lines of sight, this correction is a small blueshift: $\Delta v_{\rm gsr} \equiv v_{\rm gsr} - v_{\rm lsr} = -23$~\kms\ for the \mrk\ sight line and $-31$~\kms\ for \pks. For a cylindrical outflow, the GSR velocity is a projection of a vertical wind velocity onto the line of sight, implying that the GSR and wind velocities are related by $v_{\rm w} \equiv v_{\rm z} = v_{\rm gsr}/\sin{|b|}$. For this outflow geometry, a positive GSR velocity indicates an outflowing wind, while a negative GSR velocity indicates infalling gas. 

The relationship between GSR velocity and outflow velocity for a conic outflow is not as straightforward. For this outflow geometry, the observed velocity is the projection of a purely radial outflow onto the line of sight, and the observed and outflow velocities are related by
\begin{equation}
\frac{v_{\rm gsr}}{v_{\rm w}} = \frac{\rho}{r}\left(1 - \cos^2{b}\,\cos^2{l}\right) \mp \cos{b}\,\cos{l}\left[1 - \frac{\rho^2}{r^2}\left(1 - \cos^2{b}\,\cos^2{l}\right)\right],
\label{eqn:conic}
\end{equation}
where $\rho$ is the distance along the line of sight, $R_{\Sun}$ is the Solar radius, and $r$ is the distance from the GC ($r^2 \equiv \rho^2 - 2\rho R_{\Sun}\cos{b}\,\cos{l} + R_{\Sun}^2$). The minus sign is used when $\rho \leq R_{\Sun}\sec{b}\,\sec{l}$ (i.e., the absorber is closer than the plane which passes through the GC and whose normal is the line passing through the GC and the Sun) and the plus sign is used otherwise.  One consequence of Equation~\ref{eqn:conic} is that a conic outflow is observed to have a negative line-of-sight velocity when $\rho < \rho_0$, the distance at which $v_{\rm gsr}/v_{\rm w} = 0$. Numerically solving Equation~\ref{eqn:conic} using the coordinates from Table~\ref{tab:targets} and $R_{\Sun} = 8.5$~kpc, we find $\rho_0 = 4.8$~kpc and 7.1~kpc for the \mrk\ and \pks\ sight lines, respectively. In both cases an object at $\rho_0$ is more distant than the comparison stars, which lie at distances of $\rho_* \leq 3.6$~kpc for \mrk\ and $\rho_* \leq 2.4$~kpc for \pks, but closer than the GC.

Since the HV absorbers are more distant than the comparison stars, an outflowing conic wind can produce a negative observed velocity only if it intersects our line of sight at $\rho_* < \rho < \rho_0$. These distances correspond to opening angles for the conic outflow of $110\degr \leq \alpha \leq 130\degr$ and $75\degr \leq \alpha \leq 160\degr$ in the \mrk\ and \pks\ sight lines, respectively, where 
\begin{equation}
\tan{\left(\frac{\alpha}{2}\right)} = \sqrt{\frac{R_{\Sun}^2 - 2\rho R_{\Sun} \cos{b}\,\cos{l} + \rho^2\cos^2{b}}{\rho^2(1 - \cos^2{b})}}.
\label{eqn:ang}
\end{equation}
The maximum opening angle for biconic outflows observed by \citet{heckman90} in nearby starburst galaxies was $78\degr$. Therefore, it is unlikely that the wind from our GC would have an opening angle significantly larger than $78\degr$, and thus unlikely that any outflowing gas could be blueshifted along these sight lines.

\subsection{Kinematic Models}
\label{kinmod}

We adopt the biconic+cylindrical wind model of \citet{bland-hawthorn03} as our assumed outflow geometry. We have assumed that the HV absorbers in the \mrk\ sight line are entrained in the cylindrical region of the wind since the \mrk\ sight line probes higher Galactic latitudes than the \pks\ sight line. This assumption yields outflow velocities of $v_{\rm w} = 30\pm10$~\kms\ and $90\pm15$~\kms\ for absorbers N1 and N2, respectively, using the conversion $v_{\rm w} = v_{\rm gsr}/\sin{|b|}$. These absorbers were used to estimate the radius of the cylindrical outflow by assuming that absorber N2 originates in material on the near side of the cylinder and absorber N1 is caused by material on the far side (see Figure~\ref{fig:geometry}). Material that intersects the line of sight on the far side of the cylinder is higher above the plane than material intersecting the line of sight on the near side. Thus, the width of the cylinder can be estimated by imposing the condition that the lower-velocity absorber on the far side of the cylinder reaches the same maximum height as the higher-velocity absorber on the near side of the cylinder.

We have used the Galactic gravitational potential models of \citet{allen91} and \citet*{sakamoto03} to calculate the cylindrical radius, $R_{\rm cyl}$, and maximum height, $z_{\rm max}$, that satisfy this constraint. First, the heights at which the \mrk\ sight line intersects a cylinder with radius $R_{\rm cyl}$ was calculated, followed by the value of the gravitational potential at these positions. Then the height was increased while $R_{\rm cyl}$ was held constant until a value of $z_{\rm max}$ was found where the gravitational potential differed from its initial value by $v_{\rm w}^2/2$. The value of $R_{\rm cyl}$ was then varied and the process repeated until $z_{\rm max}$ converged to the same value for both absorbers. Averaging the results from all of the potential models yields $R_{\rm cyl} = 1.6\pm0.1$~kpc and $z_{\rm max} = 12.6\pm0.1$~kpc. Using these values, the observed heights of absorbers N1 and N2 are $z = 12.5\pm0.1$~kpc and $11.5\pm0.1$~kpc, respectively (i.e., these absorbers are observed near apoGalacticon).

Analytic gravitational potential models \citep[\citealp*{paczynski90, allen91, johnston95};][with parameters taken from \citealt*{dinescu99}]{sakamoto03} were used to predict the escape velocity at these positions. Averaging the results from all of the models yields an escape velocity of $v_{\rm esc} = 520\pm80$~\kms\ at the location of absorber N1 and $v_{\rm esc} = 530\pm80$~\kms\ at the location of absorber N2. These results indicate that absorber N1 has $\sim 5$\% of the velocity necessary to escape from its current position in the Galactic gravitational potential and absorber N2 has $\sim 15$\% of the necessary velocity. Thus, the \mrk\ absorbers are not part of an unbound Galactic wind.

Absorber S1 has a negative line of sight velocity that complicates the interpretation of the \pks\ absorbers in the context of this model. The ROSAT 1.5~keV image of the GC requires that the Galactic wind must be conical to a height of $|z| \geq 2$~kpc and the HV absorbers in the \mrk\ sight line indicate that the cylindrical region of the Galactic wind has a radius of $R_{\rm cyl} = 1.6$~kpc. Combining these results implies that the conic wind region has an opening angle of $\alpha \lesssim 80\degr$. These arguments, in conjunction with the opening angles for which a conic outflow can produce a negative line of sight velocity (\S\,\ref{outflow}), imply that absorber S1 can be explained by material entrained in the near side of a conic outflow with an opening angle of $75\degr \leq \alpha \leq  80\degr$. From Equations~\ref{eqn:conic} and \ref{eqn:ang}, the GSR and wind velocities for a conic outflow with $75\degr \leq \alpha \leq  80\degr$ are related by $-0.07 \leq v_{\rm gsr}/v_{\rm w} \leq 0$. Thus the GSR velocity of absorber S1 ($-136\pm12$~\kms) corresponds to a wind velocity of $v_{\rm w} \gtrsim 2000$~\kms\ if it represents material entrained in a conic outflow! 

This geometry is unlikely since a conic outflow with an opening angle of $\sim 80\degr$ would intercept the \pks\ sight line at a cylindrical radius of $\sim 3$~kpc from the GC, roughly twice the value predicted by the HV absorbers in the \mrk\ sight line. However, even if the value of $R_{\rm cyl}$ derived from the \mrk\ absorbers is incorrect, or if the wind is asymmetric about the mid-plane, absorber S2 does not yield a similarly high estimate of the wind velocity, regardless of whether we assume that it is entrained in a conic or a cylindrical outflow. If absorber S2 is material entrained in the far side of a conic outflow with an opening angle of $\sim 80\degr$ then the absorbing gas is 21.2~kpc from the Sun at a radius of 9.6~kpc from the GC and a height of 11.4~kpc from the plane. If the absorbing gas were truly at this position then the model of \citet{bland-hawthorn03} would break down because it requires the conversion from a conic to a cylindrical outflow to occur at $R_{\rm cyl} < R_{\Sun}$. It is also hard to imagine a conical outflow remaining coherent to $R > R_{\Sun}$. Ignoring these considerations, the GSR velocity of material in the \pks\ sight line entrained in a conic outflow and located 21.2~kpc from the Sun is related to the outflow velocity by $v_{\rm gsr}/v_{\rm w} = 0.95$ (Equation \ref{eqn:conic}), implying a wind velocity of $\sim 140$~\kms\ for absorber S2 ($v_{\rm gsr} = 137\pm10$~\kms). This wind velocity is clearly inconsistent with the estimate of $v_{\rm w} \gtrsim 2000$~\kms\ derived from absorber S1.

Alternately, the analysis of absorber S1 indicates that absorber S2 could be material on the far side of a cylindrical outflow with $R_{\rm cyl} \sim 3$~kpc. For this geometry, absorber S1 is caused by gas entrained in the near side of a central conic outflow and absorber S2 is located in the cylindrical region of the wind after the transition from the central conic region has occurred. This scenario requires absorber S2 to be located at a height of $|z| \sim7$~kpc and implies a wind velocity of $v_{\rm w} = 250\pm20$~\kms\ ($v_{\rm gsr}/v_{\rm w} = \sin{|b|}$). Again, this wind velocity is inconsistent with the estimate of $v_{\rm w} \gtrsim 2000$~\kms\ derived from absorber S1. So, while fortuitous geometry can create modest radial velocities from very high total wind speeds, these high speeds are not consistent with the observed radial velocities of the other HV absorbers (i.e., it is unlikely that all four absorbers are moving nearly in the plane of the sky).

A cylindrical model is the {\em only} plausible outflow model that consistently explains the observed radial velocities of all the HV absorbers. Therefore, our preferred wind model is that of a central conic outflow with an opening angle of $\alpha \lesssim 80\degr$ that evolves into a cylindrical outflow at a radius of $R_{\rm cyl} \approx 1.6$~kpc, as indicated by the HV absorbers in the \mrk\ sight line. In this model the transition from a conic to a cylindrical outflow occurs at $|z| \sim 2$~kpc as shown in Figure~\ref{fig:geometry}. Since the \pks\ sight line intersects the surface of a cylinder with a $1.6\pm0.1$~kpc radius at $z = -4.8\pm0.1$~kpc and $-5.9\pm0.1$~kpc, absorbers S1 and S2 are both located in the cylindrical region of the outflow. Consequently, the GSR velocities of absorbers S1 and S2 correspond to wind velocities of $v_{\rm w} = -250\pm20$ and $+250\pm20$~\kms, respectively. Thus, absorber S1 represents gas falling toward the GC. Again, this is not consistent with an unbound Galactic wind, but it is consistent with a bound Galactic ``fountain''.

We have used the same gravitational potential models as in the \mrk\ sight line to calculate the escape velocity and maximum height for absorber S2. If it is located on the far side of the cylindrical outflow then the escape velocity of absorber S2 is $v_{\rm esc} = 560\pm80$~\kms. The derived wind velocity for absorber S2 is  $250\pm20$~\kms, which implies that this absorber will reach a maximum height of $|z_{\rm max}| = 12.8\pm1.0$~kpc. This height is remarkably close to the value of $z_{\rm max} = 12.6\pm0.1$~kpc obtained for absorbers N1 and N2. The derived wind speed for absorber S2 is significantly larger than the speeds found for the \mrk\ absorbers, but it is still only $\sim 45\%$ of the velocity necessary for this absorber to escape from its inferred position in the Galaxy's potential well\footnote{If absorber S2 is located on the near side of the cylindrical outflow instead, we find an escape velocity of $v_{\rm esc} = 580\pm70$~\kms\ and a maximum height of $|z_{\rm max}| = 10.6\pm0.8$~kpc. At this position the derived wind velocity for absorber S2 is also $\sim 45\%$ of the escape velocity.}. 

We have calculated the maximum height of absorber S1 by assuming that it is bound to the Galaxy and represents material that was ejected from the GC in an outburst previous to the one that ejected the other HV absorbers. Under this assumption, absorber S1 has had time to reach its maximum height and fall back toward the plane. If absorber S1 is located on the near side of the cylindrical outflow then its maximum height was $|z_{\rm max}| = 10.6\pm0.8$~kpc, and if it is located on the far side its maximum height was $|z_{\rm max}| = 12.6\pm1.0$~kpc. These maximum heights are also remarkably consistent with the values obtained for the \mrk\ absorbers.

The observed positions and velocities of all of the HV absorbers in our preferred outflow model are summarized in Table~\ref{tab:velpos}, as well as the escape velocity at an absorber's observed position and the maximum height above the plane of its assumed orbit. Table~\ref{tab:velpos} also lists an estimate for the time since the HV absorbers were ejected from the GC, assuming a cylindrical outflow model. The lower limit on the ejection time was calculated by assuming that an absorber was ejected with some initial wind speed and has since decelerated to its current velocity solely under the influence of gravity (i.e., no drag or mass loading). The upper limit to the ejection time assumes that an absorber was ejected at its current wind speed and has travelled to its observed location at a constant velocity. No upper limit was calculated for absorber S1 because it has already reached apoGalacticon and is now falling toward the disk. The method that we have used to calculate upper limits on the ejection timescale breaks down under these circumstances since objects spend most of their orbital period near apoGalacticon. These timescales are discussed in more detail in \S\,\ref{photomod}.

The HV absorbers in both sight lines support the assertion that our Galaxy does not produce an unbound Galactic wind. Moreover, the detection of gas falling toward the GC argues that it produces a Galactic fountain instead. Table~\ref{tab:velpos} shows that the \mrk\ absorbers that probe the northern outflow and the \pks\ absorbers that probe the southern outflow predict that gas on both sides of the GC will reach similar maximum heights. This consistency in $|z_{\rm max}|$ suggests that all of the absorbers were ejected from the GC with the same initial velocity of $\sim800$~\kms\ and is evidence for a GC outflow origin for the HV absorbers. However, even if the wind we have detected emanates from the disk between the comparison stars and the GC, it will not escape the Galaxy's gravitational potential, assuming similar wind geometry to that described above.

It is also possible that the HV absorbers are associated with undetected star-forming regions on the far side of the GC rather than with star formation from the GC itself, or that the HV absorbers are typical highly-ionized HVCs on the far side of the GC that have no direct association with sites of massive star formation. However, even if the HV absorbers do lie well beyond the GC, they are still almost certainly bound to the Galaxy. Assuming that they are moving purely vertically, the \mrk\ absorbers are bound to the Galaxy as long as they lie above its disk and not $\gtrsim 170$~kpc from the Sun (i.e., at $z \lesssim 140$~kpc) according to the potential models of \citet{sakamoto03}. Similarly, these models predict that the \pks\ absorber S2 is bound to the Galaxy as long as it lies within $\sim 80$~kpc of the Sun ($|z| \lesssim 45$~kpc). Clearly, the conclusion that these absorbers are not part of an unbound Galactic wind appears robust.

\citet{putman03} observed several HVCs in H$\alpha$ emission to determine their distances and found two distance solutions for each cloud depending on its location over the Galaxy's spiral arms. The far-field distances and latitudes of the sight lines observed by \citet{putman03} correspond to heights above the plane of $|z| = 2.0$--30.8~kpc, with an average value of $\langle|z|\rangle = 9.9\pm1.5$~kpc. Our HV absorbers are observed to be at heights of $|z| = 5$--12~kpc and are predicted to reach maximum heights of $12\pm1$~kpc (see Table~\ref{tab:velpos}), all of which are consistent with the HVC heights found by \citet{putman03}. Thus, the HV absorbers are not only kinematically similar to HVCs but they have a similar distribution of heights above the plane as well. Furthermore, the $z$-heights for the HVC population studied by \citet{putman03} imply that even if the HV absorbers are associated with HVCs on the far side of the GC, they should certainly be close enough to be bound to the Galaxy (i.e., at $|z| \lesssim 45$~kpc).

\section{Ionization Conditions}
\label{ionization}

The relative strength of low- and high-ionization absorption gives us insight into the ionization conditions and metallicity of the HV absorbers. Table~\ref{tab:coldens} summarizes the measured column densities of the HV absorbers in several ionic species. The FUSE bandpass covers \HI\ Lyman series lines from Ly$\beta$ down to the Lyman limit, and we have used Ly$\zeta$ through Ly$\iota$ to constrain \NHI\ for the HV absorbers. These Lyman series lines were chosen because the expected H$_2$ absorption in the AGN sight lines is weak \citep{gillmon06}, there are no \ion{O}{1} lines at the HV absorber velocities, and the low-velocity absorption is not as saturated as in the higher-order Lyman lines.  \HI\ column densities or 3$\sigma$ upper limits are listed in Table~\ref{tab:coldens} for all of the HV absorbers.

Comparing the absorbers in the \mrk\ and \pks\ sight lines, we find that they are detected with similar column densities in \CIII, \SiII, and \OVI. Interestingly, absorber N1 is only detected in low and moderate ionization species (\CIII, \CIV, \SiII, \SiIV), while absorber S1 is detected only in \NV\ and \OVI, although the wavelengths of \CIV\ and \SiIV\ are outside of the available wavelength coverage for this sight line (see Table~\ref{tab:obslog}). Absorbers N2 and S2 have nearly identical ratios of \NSiII/\NCIII\ and \NNV/\NOVI: absorber N2 has $\log{(\NSiII/\NCIII)} = -1.15\pm0.16$ and absorber S2 has $\log{(\NSiII/\NCIII)} = -1.19\pm0.26$, while both absorbers have $\log{(\NNV/\NOVI)} < -0.39$. Moreover, the value of $\log{(\NSiII/\NCIII)} = -0.76\pm0.40$ in absorber N1 agrees with the values found for N2 and S2 within the combined errors, and the upper limits on $\log{(\NNV/\NOVI)}$ found in absorbers N2 and S2 agree with the value of $-0.57\pm0.80$ found in absorber S1. However, absorber S2 has a larger proportion of low ions to high ions than absorber N2, with the ratio of \NSiII\ or \NCIII\ to \NNV\ or \NOVI\ greater in absorber S2 by 0.3--0.4~dex in all cases. Excepting that absorber N1 lacks a highly-ionized component, the absence of solid \NHI\ values for these absorbers and the absence of \SiIV\ and \CIV\ data for \pks\ make more detailed studies or comparisons currently impossible.

\subsection{Photoionization Models}
\label{photomod}

Since we have \CIV\ and \SiIV\ data for the \mrk\ absorbers, we compare them to highly-ionized HVCs whose ionization states have been studied in detail. Table~\ref{tab:hvc} shows \HI\ column densities, logarithmic column density ratios, and metallicities for all highly-ionized HVCs with metallicities inferred from photoionization modelling that have been detected in \OVI\ and either \SiIV, \CIV, or \NV\ \citep*{sembach99,fox04,fox05,collins03,collins04,ganguly05} and absorber N2. The HVCs in Table~\ref{tab:hvc} are arranged in order of decreasing \NHI. There is no strong trend in metallicity as a function of \NHI\ due to the large errors in the model metallicities.

The relative strengths of the low- and high-ionization species cannot be reproduced by a single-phase model for all of the HVCs in Table~\ref{tab:hvc}. The low-ionization species \SiII, \SiIII, \CII, \CIII, and in some cases \SiIV\ and \CIV, can be explained by a single photoionization model in which the clouds are irradiated by the extragalactic ionizing background (some models also include a contribution from the Galactic radiation field), but the best-fit model that reproduces the low ions underpredicts the \OVI\ column density by at least an order of magnitude regardless of the details of the model radiation field. Thus, the \OVI\ and \NV\ (and usually \SiIV\ and \CIV) are underpredicted by the photoionization models and must be produced by collisional ionization processes.

Absorber N2 has the second highest values of \NCIV/\NSiIV\ and \NOVI/\NCIV\ in Table~\ref{tab:hvc}, ignoring upper and lower limits. Furthermore, the value of \NSiII/\NSiIV\ for absorber N2 is lower than that found in all but one HVC and the limit on \NCII/\NCIV\ is more stringent than all but one of the other upper limits and lower than all of the \NCII/\NCIV\ detections. These results suggest that absorber N2 is enhanced in \CIV, \SiIV, and \OVI\ relative to most highly-ionized HVCs. This enhancement suggests that collisional ionization processes are more important in absorber N2 than in typical highly-ionized HVCs. A search of this sight line for X-ray absorption from \ion{O}{7} and \ion{O}{8} would clarify this issue.

The HVCs in Table~\ref{tab:hvc} have all been found to have subsolar metallicities in the range $Z = 0.06$--0.5\,$Z_{\Sun}$ \citep{fox05,collins03,collins04,ganguly05}. The HVC that most closely resembles absorber N2 is the $-270$~\kms\ HVC toward PKS~2155--304 \citep{collins04}. This absorber has the smallest \HI\ column of all the HVCs in Table~\ref{tab:hvc} with $\log{\NHI} = 15.23^{+0.38}_{-0.22}$ and all of its metal-line column densities agree with those in absorber N2 to within the combined errors except for \NCIV, which is higher by 0.5 dex in the HVC than in absorber N2. \citet{collins04} find a metallicity of 12\% Solar for the photoionization model that best reproduces the observed metal-line abundances of this HVC with an assumed $\NHI\ = 10^{15.01}~{\rm cm^{-2}}$ (the smallest allowable \NHI\ in their 1$\sigma$ column density range). Since absorber N2 has an upper limit of $\NHI\ < 10^{14.85}~{\rm cm^{-2}}$, we use the HVC photoionization model to set a lower limit on the metallicity of the HV absorbers of $Z > 0.12~Z_{\Sun}$.

The only HV absorber in which \HI\ was detected in high-order Lyman series lines is absorber S2 (see Table~\ref{tab:coldens}). The HVC in Table~\ref{tab:hvc} with \NHI\ closest to the value for absorber S2 is the $-140$~\kms\ HVC toward PKS~2155--304 \citep{collins04}, which has $\log{\NHI} = 16.37^{+0.22}_{-0.14}$. However, all of the metal-line column densities in this HVC are higher than those detected in absorber S2 by 0.3--0.6 dex. \citet{collins04} find a metallicity of 20\% Solar for the photionization model that best reproduces the metal-line abundances of this HVC at an assumed $\NHI\ = 10^{16.59}~{\rm cm^{-2}}$ (the largest allowable \NHI\ in their 1$\sigma$ column density range). Since absorber S2 has a best-fit value of $\NHI\ = 10^{16.66\pm0.83}~{\rm cm^{-2}}$ and its metal-line column densities are smaller than those of the HVC, we use the HVC photoionization model to set an upper limit on the metallicity of the HV absorbers of $Z < 0.20~Z_{\Sun}$. 

In summary, we have attempted to use the photoionization models of \citet{collins04} for two highly-ionized HVCs toward PKS~2155--304 to constrain the metallicity of the HV absorbers in the range 12--20\% Solar. While this metallicity range is lower than would be expected for a wind emanating from the GC in the recent past, this result is very uncertain due to the uncertain ionization state of these absorbers. For the best-detected HV absorbers, the \NOVI/\NCIV\ ratio is higher than in other well studied highly-ionized HVCs, as might be expected for a still more highly-ionized GC wind. This expectation is due to the probable presence of a hard ionizing spectrum emerging from near the supermassive black hole at the GC, as well as the probability that much of the gas in these absorbers is collisionally ionized by fast shocks \citep{bland-hawthorn03}. In this case the dominant oxygen ions could be \ion{O}{7} and \ion{O}{8}, as detected in several sight lines at $z \approx 0$ with {\em Chandra} \citep{nicastro02}. If this is the case, by using the strength of the \OVI\ lines we have significantly underestimated the metallicity. Long spectroscopic observations of the \pks\ sight line could settle this issue by detecting these high ions in X-ray absorption.

Our low metallicity estimates do not necessarily preclude a nuclear starburst origin for the HV absorbers. \citet{cappi99} found that 2--10~keV {\em BeppoSax} spectra of the nearby starburst galaxies NGC~253 and M~82 are best fitted by a thermal emission model with $kT \sim 6$--9~keV and a metallicity of 0.1--0.3 Solar. However, this metallicity estimate is highly uncertain and depends strongly on the choice of spectral model (e.g., number of thermal components, contribution from power-law component) and \HI\ absorbing column densities, as well as the instrument with which the data were obtained \citep{dahlem00}. By fitting the same {\em BeppoSax} data as \citet{cappi99} with a different spectral model, \citet{dahlem00} find a best-fit metallicity for NGC~253 of $16\pm2$\% Solar and two almost equally good fits to the M~82 data, one with a metallicity of 13\% Solar and one with an abundance 17 times Solar! Similarly, after analyzing all available X-ray data on NGC~253 and M~82, \citet*{weaver00} concluded that subsolar metallicities are not required. Surprisingly, the superior resolution of {\em Chandra} does not improve matters; \citet{strickland02} find that best-fit multiphase thermal spectral models of {\em Chandra} ACIS spectra of NGC~253 have iron and oxygen abundances $< 10$\% Solar. While these metallicity estimates are highly uncertain, it is still interesting to note that the winds of nearby starburst galaxies have been found to have metallicities comparable to those of our HV absorbers.

\subsection{Collisional Ionization Mechanisms}
\label{collmod}

\citet{fox05} studied the high-ion column density ratios for all of the HVCs in Table~\ref{tab:hvc} and determined that the high-ion column density ratios in highly-ionized HVCs are not appreciably different than the values measured in the Galactic halo by \citet{zsargo03}, suggesting that similar ionization processes are present in both environments \citep[see, e.g.,][]{indebetouw04a}. Several collisional ionization mechanisms have been proposed, including collisional ionization equilibrium \citep{sutherland93}, hot radiatively cooling gas \citep{edgar86}, the conductive interfaces that arise when hot and cold gas come into contact in the presence of a magnetic field \citep*{borkowski90}, shock ionization as an absorber passes supersonically through its surrounding medium \citep{dopita96}, and turbulent mixing layers near an interface of hot and cold gas \citep*{slavin93}. \citet{fox05} found that conductive interfaces can reproduce the \NCIV/\NOVI\ and \NNV/\NOVI\ ratios in 11 of the 12 highly-ionized HVCs in their study, but they have problems reproducing the observed values of \NSiIV/\NOVI. Due to the large error bars on \NOVI\ for our HV absorbers (see Table~\ref{tab:coldens}), the values of \NCIV/\NOVI\ and \NSiIV/\NOVI\ for absorber N2 and \NNV/\NOVI\ for absorber S1 can simultaneously be explained by any of the collisional ionization mechanisms above with the exception of collisional ionization equilibrium \citep{fox04,fox05}. However, all of the plausible mechanisms suggest that the high ions in the HV absorbers reside in cloud boundaries, forming a collisionally ionized skin around the photoionized bulk of the cloud. Thus, we expect that the \CIV, \SiIV, \NV\ and \OVI\ in the HV absorbers are found at the absorber boundaries, with the photoionized low ions in the interior.

\subsection{Low Velocity Absorption}
\label{lowvel}

It is interesting to note that the \OVI\ absorption is stronger in the AGN spectra than in the comparison stars at all velocities except for $100 \lesssim v_{\rm lsr} \lesssim 150$~\kms\ in the \pks\ sight line (see Figure~\ref{fig:datafit}). For all other ions in Figure~\ref{fig:datafit} (with the exception of \NV\ in the \mrk\ sight line) the comparison star absorption has comparable strength to the AGN absorption at low velocities. There is a considerable amount of small- and large-scale structure in the distribution of \OVI\ absorbing gas in the Galactic halo, as evidenced by large differences in \OVI\ absorption on scales of $< 1\degr$ to $180\degr$ \citep{savage03,sembach03}. The \mrk\ sight line is separated by $15\degr$ and $13\degr$, respectively, from the HD~121968 and HD~125924 sight lines (see Figure~\ref{fig:posn}) and is higher in total \OVI\ column density by 0.7 and 0.6~dex. The angular separation and column density variations are smaller for the \pks\ sight line (Figure~\ref{fig:posn}), which is separated by $9\degr$ and $10\degr$ and is higher in total \OVI\ column density by 0.3 and 0.4~dex from the HD~187311 and HD~191466 sight lines, respectively. These variations in \OVI\ absorbing column are all considerably larger than the average variations found by \citet{savage03} in sight lines with similar angular separations. Thus, the AGN sight lines also probe low velocity regions of the halo that are significantly more highly-ionized than the comparison star sight lines. These lower velocity, highly-ionized regions must be located beyond the comparison stars ($\rho > 3.6$~kpc) and are most likely in the lower halo between us and the GC.

\section{Conclusions}
\label{conclusion}

We have used FUSE and HST spectra of the UV-bright AGN \mrk\ and \pks\ to detect absorption that most probably arises from a nuclear wind emanating from the center of our Galaxy. Spectra of four comparison stars, two for each AGN, were used to identify and remove foreground velocity components from the absorption profiles of \OVI\ $\lambda1032$, \NV\ $\lambda\lambda1239, 1243$, \CII\ $\lambda1036$, \CII\ $\lambda1335$, \CIII\ $\lambda977$, \CIV\ $\lambda\lambda1548, 1551$, \SiII\ $\lambda1260$, \SiIII\ $\lambda1207$, and \SiIV\ $\lambda\lambda1394, 1403$ in the AGN spectra. There is a large amount of overlap in the high-velocity (HV) absorption velocities from ion to ion, indicating that these ions coexist spatially along the AGN lines of sight. Two HV absorbers are detected toward \mrk: absorber N1 is detected in \CIII, \CIV, \SiII, and \SiIV\ at an average LSR velocity of $+45\pm7$~\kms\ and absorber N2 is detected in \CIII, \CIV, \SiII, \SiIV, and \OVI\ at an average LSR velocity of $+94\pm12$~\kms. We also detect two HV absorbers in the \pks\ sight line, absorber S1 in \NV\ and \OVI\ at a velocity of $-105\pm12$~\kms, and absorber S2 in \CIII, \SiII, and \OVI\ at a velocity of $+168\pm10$~\kms. 

These HV absorbers are consistent in velocity with high- and intermediate-velocity clouds seen in other sight lines. However, due to the proximity of our sight lines to $l=0\degr$ and the detection of the HV absorbers in highly-ionized species such as \CIV, \SiIV, \NV, and \OVI, we hypothesized that they were created by material entrained in an outflowing Galactic wind. While we cannot prove decisively that these absorbers are due to gas above and below the GC, this is their most likely location because: (1) a GC wind has been indirectly detected using various techniques from the mid-IR to X-rays; (2) similar winds are observed in other galaxies to emanate from above active sites of star formation, especially from nuclear starbursts \citep[as is on-going in our own GC;][]{figer02}; (3) based upon a large survey of Galactic halo \OVI\ absorbers \citep{savage03,sembach03} it is quite unusual to detect several highly-ionized HVCs without corresponding \HI\ 21~cm emission along individual sight lines. The \pks\ \OVI\ absorption line is one of the two strongest detected in this survey. An association between these absorbers and the GC would explain these anomalies; (4) the assumption of a GC point of origin for these absorbers leads to the conclusion that they will rise to a similar height above the mid-plane ($12\pm1$~kpc) on both sides of the plane, as might be expected if they were all due to a single nuclear ``explosion'' $\geq 50$~Myr ago. 

However, based upon a comparison with other highly-ionized HVCs we crudely estimate a metallicity of these absorbers of only $\sim 10$--20\% Solar, lower than expected for GC ejecta. We consider this estimate quite uncertain due to the unknown ionization state of these absorbers. GC ejecta might be expected to be very highly ionized either because of collisional ionization in fast shocks and/or photoionization from a hard ionizing spectrum coming from the black hole in the GC. A long spectroscopic observation of \pks\ with {\em Chandra} has the potential to determine more precisely the ionization state and thus the metallicity of the absorbing gas. Interestingly, X-ray spectra of the nearby starburst galaxies NGC~253 and M~82 yield highly uncertain metallicity estimates that are comparable to our crude estimate of the metallicity of our HV absorbers \citep{cappi99,weaver00,dahlem00,strickland02}.

We adopt the outflow geometry of \citet{bland-hawthorn03}, in which a central biconic outflow evolves into a cylindrical wind at large distances from the Galactic plane. The only outflow model that consistently explains the observed radial velocities of the HV absorbers requires that they represent gas entrained in the cylindrical region of this wind. Absorbers N1 and N2 have outflow velocities of $v_{\rm w} = +30$ and +90~\kms\ and are located 12.5 and 11.5~kpc above the plane, respectively, and absorbers S1 and S2 have outflow velocities of $v_{\rm w} = -250$ and +250~\kms\ and are located 4.8 and 5.9~kpc below the plane, respectively. We have used several analytic gravitational potential models of the Milky Way \citep{paczynski90, allen91, johnston95, sakamoto03} to calculate the escape velocity at these locations. These models predict escape velocities of $\gtrsim 550$~\kms\ at the locations of each of the HV absorbers. Since each HV absorber has $\leq 45$\% of the velocity necessary to escape from its current position in the Galactic gravitational potential, our absorbers are consistent with a bound Galactic ``fountain''. This conclusion is strengthened by the detection of gas falling toward the GC in absorber S1 and by the consistent derived maximum height of $|z_{max}| \approx 12$~kpc for all absorbers.

The HV absorbers are kinematically similar to HVCs and are located at similar heights above the plane \citep{putman03}. \citet{keeney05} found a similar result for the nearby luminous starburst galaxy NGC~3067. Using the background quasar 3C~232 to probe the halo of NGC~3067 near the minor axis 11$\;h_{70}^{-1}$~kpc from the galactic plane, they found a consistent velocity structure for various metal species and ionization states (e.g., \ion{Na}{1}, \ion{Ca}{2}, \ion{Mg}{1}, \SiIV, \CIV). \citet{keeney05} suggested that these absorbers are analogous to Galactic HVCs and found them to have comparable \HI\ column densities, kinematics, metallicities, spin temperatures, and inferred sizes. Furthermore, they found no evidence that any halo gas along the line of sight to 3C~232 is escaping NGC~3067, despite its modest starburst. Thus, the overall physical picture of the 3C~232/NGC~3067 system is also that of a galactic fountain and not an outflowing starburst wind.

Regardless of whether or not the HV absorbers represent Galactic fountain material, they are useful analogs for absorption line studies of starburst winds in general. Typical outflow velocities of 400--1000~\kms\ are found in absorption line studies of starburst winds \citep{heckman00}. These studies usually use the stellar continuum of the starburst region itself as the background source, which introduces an ambiguity in the location of the absorbing gas relative to the starburst region. Our HV absorbers are clearly bound to the Galaxy and imply an initial wind speed of $\sim 800$~\kms\ in order to reach a maximum height of $\approx 12$~kpc in the Galactic gravitational potential with no mass loading.  Thus, the high initial wind speeds measured in nearby starburst galaxies do not absolutely require that the winds escape their host galaxy's gravitational potential well either.

\citet{martin99} found that the temperature of starburst winds is nearly constant as a function of a galaxy's maximum \HI\ rotation speed, suggesting that the speed of starburst winds is independent of galaxy mass. If so, then galactic winds are more likely to escape from the shallow potentials of dwarf galaxies than those of their more massive counterparts. The results of \citet{keeney05} and those presented here suggest that it is difficult for winds to escape from luminous ($\sim L^*$) galaxies. Perhaps winds from dwarf galaxies \citep{stocke04,keeney06} were primarily responsible for the metal enrichment of the IGM.

\begin{figure}[!ht]
\begin{center}
\epsscale{1.00}
\plotone{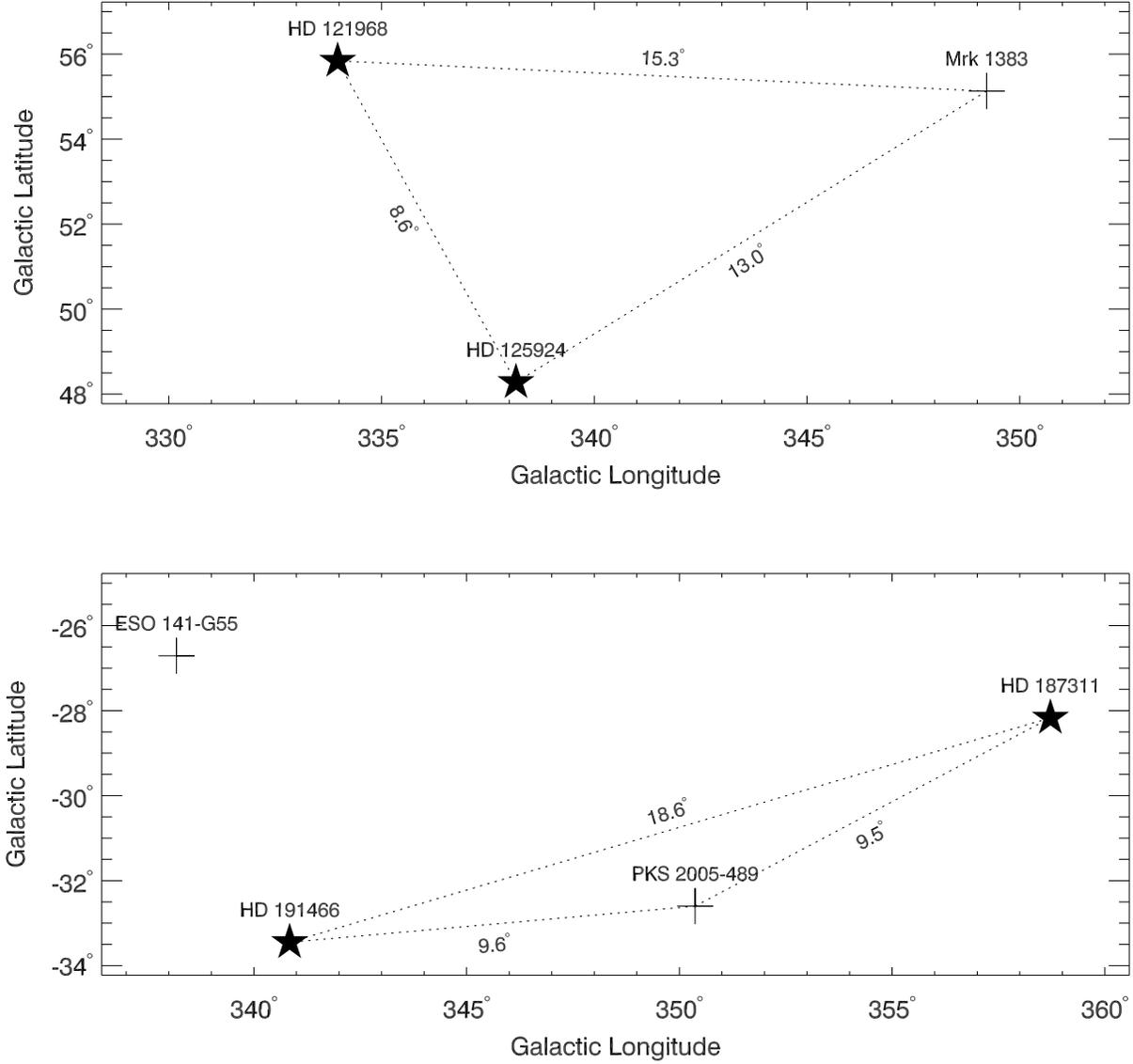}
\end{center}
\caption{Relative positions of the AGN and the comparison stars in Galactic coordinates. The angular separations between the AGN and comparison stars, in degrees, are also indicated. Positions for all objects are from Table~\ref{tab:targets}. The location of the AGN ESO~141--G55 is indicated in the bottom panel because it is close to \pks\ and HD~191466 on the sky and was found by \citet{sembach03} to have high-velocity \OVI\ absorption.
\label{fig:posn}}
\end{figure}

\begin{figure}[!ht]
\begin{center}
\epsscale{0.85}
\plotone{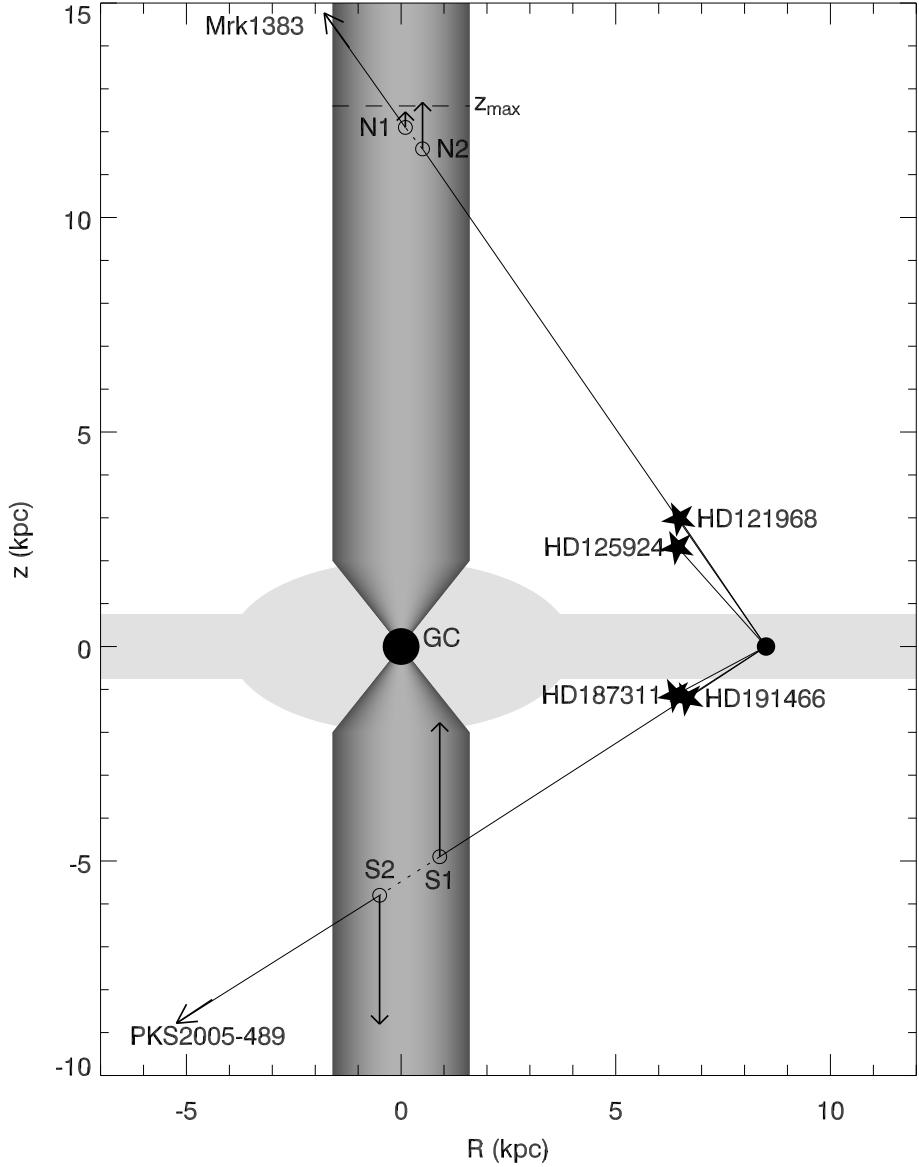}
\end{center}
\caption{A schematic diagram of the relative positions of our AGN and stellar targets. Galactic coordinates for all objects and distances to the comparison stars are from Table~\ref{tab:targets}. Our preferred wind geometry of a biconic outflow that evolves to a cylinder at $R_{\rm cyl} = 1.6$~kpc and $|z| \gtrsim 2$~kpc (see \S\,\ref{hvkin}) is represented by the dark gray region. The path lengths over which our sight lines are ``interior'' to the outflow are indicated by the dotted line segments. The relative outflow velocities and directions of the HV absorbers are shown as vectors, and the maximum height of these absorbers in the Galactic gravitational potential is indicated by a dashed line. The conical region of the outflow is consistent with the biconic outflows detected near the GC in X-ray and mid-IR emission \citep{wang02,bland-hawthorn03}.
\label{fig:geometry}}
\end{figure}

\begin{figure}[!ht]
\begin{center}
\epsscale{0.80}
\plotone{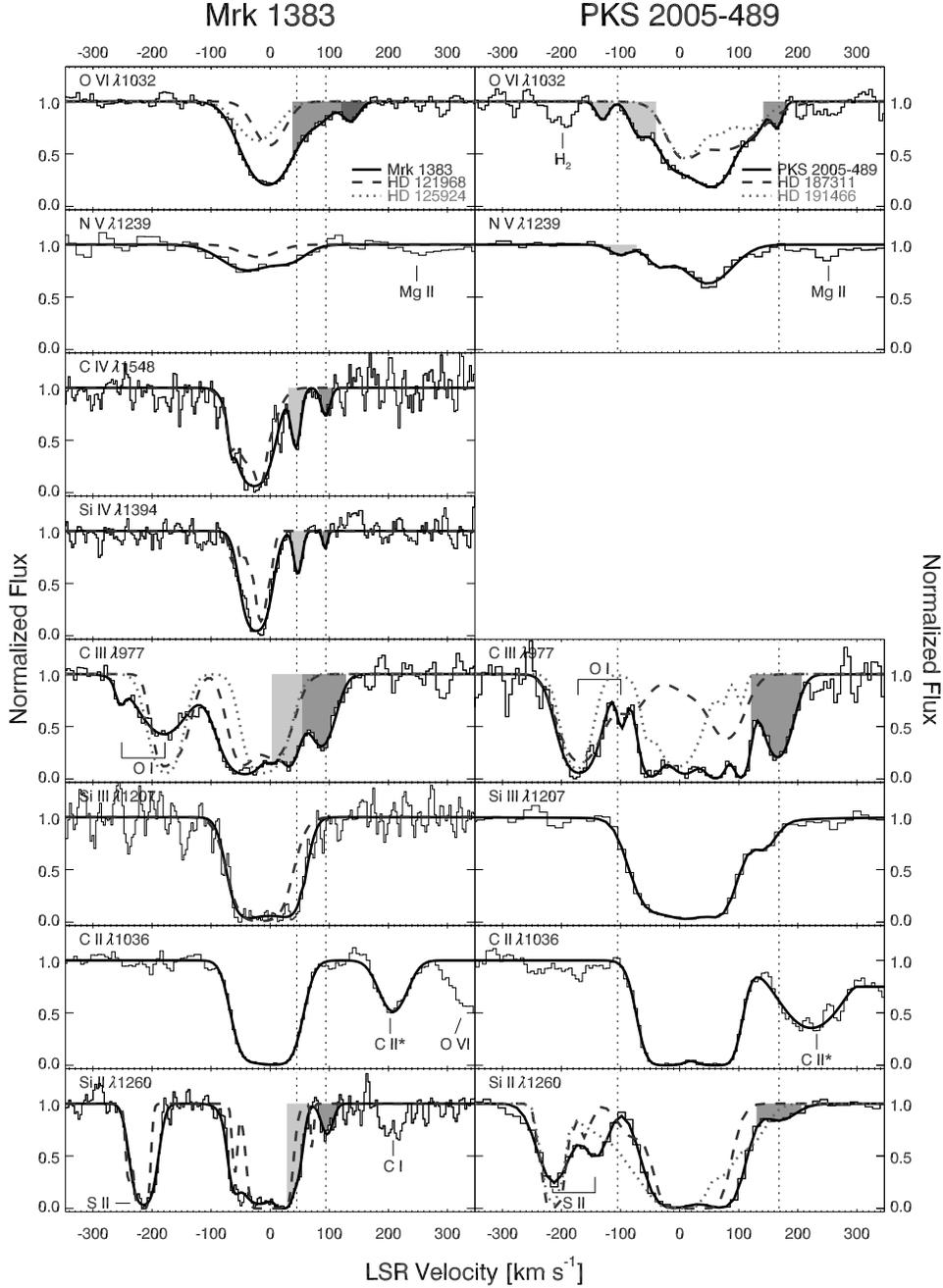}
\end{center}
\caption{Absorption line profiles ({\em histograms}) from FUSE and HST/STIS spectra of \mrk\ ({\em left}) and \pks\ ({\em right}). Best-fit Voigt profiles for the AGN spectra are shown as solid lines and best-fit Voigt profiles for the comparison star spectra (data not shown) are shown as dashed and dotted lines. The shaded regions indicate the high-velocity wind absorbers listed in Tables~\ref{tab:mrkparams} and \ref{tab:pksparams}, and the dotted vertical lines show the average velocity of these absorbers over all ions in which they were detected.
\label{fig:datafit}}
\end{figure}

For our own Galaxy's nuclear wind, there is some inconclusive evidence that these absorptions are more wide-spread than just along the two sightlines studied in detail here. The AGN ESO~141--G55 lies $\sim 12\degr$ further from $l=0\degr$ than \pks, but shows strong \OVI\ absorption at the same redshifted velocity and at a similar \NOVI\ as the \OVI\ absorber at $v_{\rm lsr} = 165\pm11$~\kms\ in the \pks\ sight line \citep{sembach03}. There are also indications of \OVI, \NV, and \CIV\ absorption at $v_{\rm lsr} \sim -65$~\kms\ in ESO~141--G55 \citep{wakker03,penton00,indebetouw04b}, and the \pks\ sight line shows \OVI\ and \NV\ absorption at similar velocities. While this is only one additional sight line, it suggests that the outflowing gas we have detected may be more widespread above and below the GC than we can determine based upon three sight lines alone. Further analysis of this highly-ionized absorption in ESO~141--G55 could provide another probe of the Galactic wind, although at distances further from the GC than those probed by \pks\ (Figure~\ref{fig:posn}). This new information could help clarify whether our HV absorbers are wide-spread Galactic fountain material or isolated HVC absorbers.

The highest ions studied herein indicate cloud velocities which do not escape the Galaxy's gravitational potential. However, it is possible that the gas associated with the Galactic wind has a temperature $\gtrsim 10^6$~K, in which case the species observed in this study represent only a trace constituent of the outflow. If the dominant wind component truly is this hot, then the kinematics of the HV absorbers would indicate that cooler material associated with the wind, whether entrained in the outflow or condensing out of the flow, does not have sufficient velocity to escape the Galaxy. In this scenario the dominant hotter, more highly-ionized gas could be moving at faster speeds, perhaps fast enough to escape. Very hot gas associated with the Galaxy has been detected by Chandra in \ion{O}{7} and \ion{O}{8} in other sight lines \citep*[e.g.,][]{rasmussen03,nicastro02,nicastro05,mckernan05}. Therefore, a search for \ion{O}{7} and \ion{O}{8} toward \mrk\ and \pks\ would test whether the dominant component of the Galactic wind is more highly-ionized than the species studied here and whether it could move to larger distances from the plane. \pks\ is bright enough to be feasibly observed with current X-ray telescopes, but a search for \ion{O}{7} and \ion{O}{8} toward \mrk\ will have to wait for the next generation of X-ray spectrographs.

The Galactic wind hypothesis for the origin of our HV absorbers is not a unique interpretation of the data. It is conceivable that the HV absorbers are typical highly-ionized HVCs with no direct association to sites of massive star formation. The only way to resolve this ambiguity is to observe additional high-latitude AGN sight lines near $l=0\degr$ to confirm the results presented here. With 10--20 times the throughput of STIS at comparable resolution, the Cosmic Origins Spectrograph will be able to observe AGN within $10\degr$ of the GC that are too faint to be observed by STIS or FUSE. Spectra of these AGN, along with spectra of appropriate comparison stars, are ideally suited for further testing the Galactic wind hypothesis.

{\acknowledgments We acknowledge support from NASA HST General Observer grant GO-09778 and NASA FUSE contract NNG05GB64G. B. A. K. acknowledges support from NASA Graduate Student Researchers Program grant NGT5-154. J. M. S. acknowledges support from NASA/LTSA grant NAG5-7262. J. T. S. acknowledges that inspiration for this research arose while reading the results of the FUSE \OVI\ survey conducted by Sembach, Wakker, Savage, Shull, et al. This work is based on observations made with the NASA/ESA {\em Hubble Space Telescope} and the NASA-CNES-CSA {\em Far Ultraviolet Spectroscopic Explorer}. The HST data were obtained at the Space Telescope Science Institute, which is operated by the Association of Universities for Research in Astronomy, Inc., under NASA contract NAS5-26555. FUSE is operated for NASA by the Johns Hopkins University under NASA contract NAS5-32985.}

%Tables
\clearpage
\begin{deluxetable}{lccccccccc}

\tabletypesize{\scriptsize}
\tablecolumns{10}
\tablewidth{0pt}

\tablecaption{Galactic Wind Targets \label{tab:targets}}

\tablehead{\colhead{} & \multicolumn{2}{c}{R.A.~~(J2000)~~Dec} & \colhead{$l$} & \colhead{$b$} & \colhead{} & \colhead{V} & \colhead{d} & \colhead{$\Delta v_{\rm lsr}$\tablenotemark{a}} \\ \colhead{Target} & \colhead{$h~~m~~s$} & \colhead{$~\degr~~\arcmin~~\arcsec$} & \colhead{(deg)} & \colhead{(deg)} & \colhead{Sp. Type} & \colhead{(mag)} & \colhead{(kpc)} & \colhead{(\kms)} & \colhead{Dist. Ref.\tablenotemark{b}}}

\startdata
\mrk      & 14~29~06.6 & $+01~17~06$ & 349.22 &  55.13 &       AGN & 17.5 & \nodata & 10 & \nodata \\
HD~121968 & 13~58~51.2 & $-02~54~52$ & 333.97 &  55.84 &      B1~V & 10.3 &     3.6 &  8 & 1     \\
HD~125924 & 14~22~43.0 & $-08~14~54$ & 338.16 &  48.28 &     B2~IV & ~9.7 &     3.1 &  8 & 1     \\
\pks      & 20~09~25.4 & $-48~49~54$ & 350.37 & -32.60 &       AGN & 15.3 & \nodata &  2 & \nodata \\
HD~187311 & 19~51~04.7 & $-41~01~24$ & 358.72 & -28.17 &      B3~V & 10.3 &     2.4 &  5 & 2     \\
HD~191466 & 20~12~58.1 & $-56~50~48$ & 340.84 & -33.45 & B5/B6~III & ~8.7 &     2.2 &  0 & 3     \\
\enddata

\tablenotetext{a}{Heliocentric to LSR velocity correction ($\Delta v_{\rm lsr} \equiv v_{\rm lsr} - v_{\rm hel}$), assuming a solar motion of 16.5~\kms\ toward $(l,b) = (53\arcdeg,25\arcdeg)$ from \citet{mihalas81}.}
\tablenotetext{b}{References for the distance to the comparison stars: (1) \citet{sembach92}; (2) \citet{hill70}; (3) luminosity distance based on an absolute magnitude extrapolated from Table 15.7 of \citet{drilling00} and Galactic dust extinction values from \citet*{schlegel98}.}

\end{deluxetable}

\clearpage
\begin{deluxetable}{lcccccc}

\tablecolumns{7}
\tablewidth{0pt}

\tablecaption{Journal of Observations \label{tab:obslog}}

\tablehead{\colhead{} & \colhead{} & \colhead{} & \colhead{} & \colhead{Coverage} & \colhead{Resolution} & \colhead{Exposure} \\ \colhead{Target} & \colhead{Instrument} & \colhead{Grating} & \colhead{Dataset ID} & \colhead{(\AA)} & \colhead{(\kms)} & \colhead{(ksec)}}

\startdata
\mrk      & FUSE     & \nodata & multiple\tablenotemark{a} & ~905--1187 & 18 & 63.3 \\
          & HST STIS & G140M   & O4EC01                    & 1194--1300 & 38 & ~7.5 \\
          & HST STIS & E140M   & multiple\tablenotemark{a} & 1140--1735 & ~9 & 19.2 \\
HD~121968 & FUSE     & \nodata & S10145                    & ~905--1187 & 18 & ~9.3 \\
          & HST STIS & E140H   & O57R02                    & 1170--1554 & ~3 & 12.8 \\
HD~125924 & FUSE     & \nodata & S10147                    & ~905--1187 & 18 & 11.0 \\
\\
\tableline
\\
\pks      & FUSE     & \nodata & multiple\tablenotemark{a} & ~905--1187 & 18 & 48.7 \\
          & FUSE     & \nodata & multiple\tablenotemark{a} & ~905--1187 & 18 & 22.8\tablenotemark{b} \\
          & HST STIS & G140M   & O4EC09                    & 1194--1300 & 38 & 11.6 \\
HD~187311 & FUSE     & \nodata & E09101                    & ~905--1187 & 18 & 49.8 \\
          & HST STIS & E140M   & O8PG05                    & 1140--1735 & ~9 & ~2.4 \\
HD~191466 & FUSE     & \nodata & E09103\tablenotemark{c}   & ~905--1187 & 18 & 14.6 \\
          & HST STIS & E140M   & O8PG07                    & 1140--1735 & ~9 & ~2.6 \\
\enddata

\tablenotetext{a}{Multiple datasets were coadded as follows: \mrk\ (FUSE) = S26701, S10148; \mrk\ (STIS) = OPG01, OPG02; \pks\ = C14903, S10738.}
\tablenotetext{b}{C\,III was observed in airglow emission in the FUSE spectrum of \pks, so night-only data with the given exposure time were used to analyze the C\,III region in this sight line.}
\tablenotetext{c}{There are no side 2 data for this observation.}

\end{deluxetable}

\clearpage
\begin{deluxetable}{lccccccccc}

\tabletypesize{\scriptsize}
\tablecolumns{10}
\tablewidth{0pt}

\tablecaption{Measurements of High Velocity Absorption Toward \mrk\ \label{tab:mrkparams}}

\tablehead{\colhead{} & \colhead{} & \multicolumn{4}{c}{\underline{Abs. N1: $\bar{v}_{\rm lsr} = 45\pm7$~\kms}} & \multicolumn{4}{c}{\underline{Abs. N2: $\bar{v}_{\rm lsr} = 94\pm12$~\kms}} \\ \colhead{} & \colhead{} & \colhead{$v_{\rm lsr}$} & \colhead{$b$} & \colhead{$W_{\lambda}$\tablenotemark{b}} & \colhead{$\log{N}$} & \colhead{$v_{\rm lsr}$} & \colhead{$b$} & \colhead{$W_{\lambda}$\tablenotemark{b}} & \colhead{$\log{N}$} \\ \colhead{Line} & \colhead{S/N\tablenotemark{a}} & \colhead{(\kms)} & \colhead{(\kms)} & \colhead{(m\AA)} & \colhead{(cm$^{-2}$)} & \colhead{(\kms)} & \colhead{(\kms)} & \colhead{(m\AA)} & \colhead{(cm$^{-2}$)}}

\startdata
C\,II~$\lambda1036.337$   & 16 & \nodata   & \nodata   & \tablenotemark{d} & \tablenotemark{d} & \nodata & \nodata & $< 12$ & $< 13.03$ \\
C\,II~$\lambda1334.532$   & 17 & \nodata   & \nodata   & \tablenotemark{d} & \tablenotemark{d} & \nodata & \nodata & $< ~8$ & $< 12.60$ \\
C\,III~$\lambda977.020$   & 12 & $33\pm12$ & $19\pm15$ & $105\pm14$ & $13.58\pm0.23$ & $87\pm13$ & $26\pm10$ & $121\pm22$ & $13.47\pm0.15$ \\
C\,IV~$\lambda1548.204$   & ~7 & $43\pm~2$ & $~9\pm~1$ & $~59\pm22$ & $13.30\pm0.04$ & $94\pm~3$ & $~9\pm~3$ & $~26\pm22$ & $12.86\pm0.09$ \\
C\,IV~$\lambda1550.781$   & ~7 & $45\pm~3$ & $11\pm~3$ & $~46\pm28$ & $13.46\pm0.08$ & $85\pm~3$ & $~2\pm~0$\tablenotemark{c} & $19\pm13$ & $13.21\pm0.19$ \\
N\,V~$\lambda1238.821$    & 19 & \nodata   & \nodata   & \tablenotemark{d} & \tablenotemark{d} & \nodata & \nodata & $< 25$ & $< 13.07$ \\
N\,V~$\lambda1242.804$    & 20 & \nodata   & \nodata   & \tablenotemark{d} & \tablenotemark{d} & \nodata & \nodata & $< 24$ & $< 13.35$ \\
O\,VI~$\lambda1031.926$\tablenotemark{e} & 18 & \nodata & \nodata & \tablenotemark{d} & \tablenotemark{d} & $82\pm22$ & $30\pm43$ & $32\pm13$ & $13.46\pm0.58$ \\
Si\,II~$\lambda1260.422$  & 10 & $44\pm~7$ & $11\pm~5$ & $~71\pm14$ & $12.82\pm0.32$ & $96\pm~4$ & $12\pm~2$ & $~28\pm16$ & $12.32\pm0.06$ \\
Si\,III~$\lambda1206.500$ & ~5 & \nodata   & \nodata   & \tablenotemark{d} & \tablenotemark{d} & \nodata & \nodata & $< 21$ & $< 12.00$\\
Si\,IV~$\lambda1393.760$  & 10 & $47\pm~2$ & $~8\pm~1$ & $~32\pm10$ & $12.67\pm0.04$ & $93\pm~3$ & $~2\pm~0$\tablenotemark{c} & $~~7\pm~6$ & $12.04\pm0.17$ \\
Si\,IV~$\lambda1402.773$  & 10 & $46\pm~5$ & $~5\pm~8$ & $~11\pm~7$ & $12.49\pm0.35$ & $89\pm~4$ & $~2\pm~0$\tablenotemark{c} & $~~5\pm~6$ & $12.14\pm0.25$ \\
\enddata

\tablenotetext{a}{The average signal-to-noise ratio per resolution element as measured by continuum deviations from a best-fit Legendre polynomial.}
\tablenotetext{b}{Equivalent width detections are quoted with $1\sigma$ errors, and upper limits are at $3\sigma$ confidence.}
\tablenotetext{c}{This $b$-value is the minimum of the allowed fitting range, so no error is available.}
\tablenotetext{d}{No measurement or upper limit possible because of blending with disk/halo absorption or a spurious nearby absorption line (see Figure~\ref{fig:datafit}).}
\tablenotetext{e}{High velocity O\,VI is also detected at $v_{\rm lsr} = 138\pm12$~\kms\ (see Figure~\ref{fig:datafit}) with $b = 15\pm19$~\kms, $\log{N ({\rm cm}^{-2})} = 13.26\pm0.41$, and $W_{\lambda} = 19\pm7$~m\AA. This absorption component was not studied further since gas at this velocity was not detected in any other ion.}

\end{deluxetable}

\clearpage
\begin{deluxetable}{lccccccccc}

\tabletypesize{\scriptsize}
\tablecolumns{10}
\tablewidth{0pt}

\tablecaption{Measurements of High Velocity Absorption Toward \pks\ \label{tab:pksparams}}

\tablehead{\colhead{} & \colhead{} & \multicolumn{4}{c}{\underline{Abs. S1: $\bar{v}_{\rm lsr} = -105\pm12$~\kms}} & \multicolumn{4}{c}{\underline{Abs. S2: $\bar{v}_{\rm lsr} = 168\pm10$~\kms}} \\ \colhead{} & \colhead{} & \colhead{$v_{\rm lsr}$} & \colhead{$b$} & \colhead{$W_{\lambda}$\tablenotemark{b}} & \colhead{$\log{N}$} & \colhead{$v_{\rm lsr}$} & \colhead{$b$} & \colhead{$W_{\lambda}$\tablenotemark{b}} & \colhead{$\log{N}$} \\ \colhead{Line} & \colhead{S/N\tablenotemark{a}} & \colhead{(\kms)} & \colhead{(\kms)} & \colhead{(m\AA)} & \colhead{(cm$^{-2}$)} & \colhead{(\kms)} & \colhead{(\kms)} & \colhead{(m\AA)} & \colhead{(cm$^{-2}$)}}

\startdata
C\,II~$\lambda1036.337$   & 13 & \nodata & \nodata & $< 14$ & $< 13.10$ & \nodata & \nodata & \tablenotemark{d} & \tablenotemark{d} \\
C\,III~$\lambda977.020$   & ~6 & \nodata & \nodata & \tablenotemark{d} & \tablenotemark{d} & $166\pm12$ & $27\pm16$ & $157\pm45$ & $13.61\pm0.23$ \\
N\,V~$\lambda1238.821$    & 33 & $-103\pm~5$ & $~5\pm~0$\tablenotemark{c} & $~17\pm12$ & $13.11\pm0.12$ & \nodata & \nodata & $< 15$ & $< 12.85$    \\
N\,V~$\lambda1242.804$    & 36 & $~-88\pm~6$ & $~5\pm~0$\tablenotemark{c} & $~~6\pm~5$ & $13.05\pm0.34$ & \nodata & \nodata & $< 13$ & $< 13.09$    \\
O\,VI~$\lambda1031.926$   & 13 & $-130\pm11$ & $10\pm19$ & $~14\pm10$ & $13.14\pm0.34$ & $165\pm11$ & $~7\pm22$ & $~18\pm11$ & $13.24\pm0.43$ \\
                          & 13 & $~-70\pm16$ & $17\pm21$ & $~33\pm17$ & $13.49\pm0.72$ \\
Si\,II~$\lambda1260.422$  & 34 & \nodata     & \nodata   & \tablenotemark{d} & \tablenotemark{d} & $170\pm~6$ & $28\pm~7$ & $~40\pm13$ & $12.44\pm0.12$ \\
Si\,III~$\lambda1206.500$ & 25 & \nodata     & \nodata   & \tablenotemark{d} & \tablenotemark{d} & \nodata    & \nodata   & \tablenotemark{d} & \tablenotemark{d} \\
\enddata

\tablenotetext{a}{The average signal-to-noise ratio per resolution element as measured by continuum deviations from a best-fit Legendre polynomial.}
\tablenotetext{b}{Equivalent width detections are quoted with $1\sigma$ errors, and upper limits are at $3\sigma$ confidence.}
\tablenotetext{c}{The $b$-value was fixed at this value because it was very poorly constrained by the fit when allowed to vary freely.}
\tablenotetext{d}{No measurement or upper limit possible because of blending with disk/halo absorption or a spurious nearby absorption line (see Figure~\ref{fig:datafit}).}

\end{deluxetable}

\clearpage
\begin{deluxetable}{lllcccc}

\tablecolumns{7}
\tablewidth{0pt}

\tablecaption{Velocities and Positions of the High Velocity Wind Absorbers \label{tab:velpos}}

\tablehead{\colhead{} & \colhead{} & \colhead{} & \multicolumn{2}{c}{\underline{\mrk\ Abs.}} & \multicolumn{2}{c}{\underline{\pks\ Abs.}} \\ \colhead{} & \colhead{} & \colhead{} & \colhead{N1} & \colhead{N2} & \colhead{S1} & \colhead{S2}}

\startdata
$v_{\rm lsr}$                    & (\kms) && $45\pm7$     & $94\pm12$    & $-105\pm12$   & $168\pm10$    \\
$v_{\rm gsr}$\tablenotemark{a}   & (\kms) && $22\pm7$     & $71\pm11$    & $-136\pm12$   & $137\pm10$    \\
$v_{\rm w}$\tablenotemark{b}     & (\kms) && $30\pm10$    & $90\pm15$    & $-250\pm20$   & $250\pm20$    \\
$v_{\rm esc}$\tablenotemark{c}   & (\kms) && $520\pm80$   & $530\pm80$   & $580\pm70$    & $560\pm80$    \\
$z_{\rm obs}$                    & (kpc)  && $12.5\pm0.1$ & $11.5\pm0.1$ & $-4.8\pm0.1$  & $-5.9\pm0.1$  \\
$z_{\rm max}$\tablenotemark{d}   & (kpc)  && $12.6\pm0.1$ & $12.6\pm0.1$ & $-10.6\pm0.8$ & $-12.8\pm1.0$ \\
$t_{\rm eject}$\tablenotemark{e} & (Myr)  && 70--400      & 55--130      & $>800$        & 20--50        \\
\enddata

\tablenotetext{a}{The velocity with respect to the Galactic standard of rest: $v_{\rm gsr} \equiv v_{\rm lsr} + (220~\kms)\,\cos{b}\,\sin{l}$.}
\tablenotetext{b}{The outflow velocity of the absorber, assuming a cylindrical outflow model.}
\tablenotetext{c}{The escape velocity at the observed absorber position.}
\tablenotetext{d}{The maximum height that the absorber will reach in the Galactic gravitational potential.}
\tablenotetext{e}{The time since an absorber was ejected, assuming a cylindrical outflow model.}

\end{deluxetable}

\clearpage
\begin{deluxetable}{lcccc}

\tablecolumns{5}
\tablewidth{0pt}

\tablecaption{Summary of Column Density Measurements \label{tab:coldens}}

\tablehead{ & \multicolumn{2}{c}{\underline{\mrk\ Abs.}} & \multicolumn{2}{c}{\underline{\pks\ Abs.}} \\ \colhead{Ion} & \colhead{N1} & \colhead{N2} & \colhead{S1} & \colhead{S2}}

\startdata
H\,I    & $< 16.05$      & $< 14.85$      & $< 16.20$      & $16.66\pm0.83$ \\
C\,II   & \nodata        & $< 12.60$      & $< 13.10$      & \nodata        \\
C\,III  & $13.58\pm0.23$ & $13.47\pm0.15$ & \nodata        & $13.61\pm0.23$ \\
C\,IV   & $13.38\pm0.08$ & $13.04\pm0.18$ & \nodata        & \nodata        \\
Si\,II  & $12.82\pm0.32$ & $12.32\pm0.06$ & \nodata        & $12.44\pm0.12$ \\
Si\,III & \nodata        & $< 12.00$      & \nodata        & \nodata        \\
Si\,IV  & $12.58\pm0.09$ & $12.09\pm0.05$ & \nodata        & \nodata        \\
N\,V    & \nodata        & $< 13.07$      & $13.08\pm0.03$ & $< 12.85$      \\
O\,VI   & \nodata        & $13.46\pm0.58$ & $13.65\pm0.80$\tablenotemark{a} & $13.24\pm0.43$ \\
\enddata

\tablecomments{All values in this table are the logarithm of the ionic column density, $\log{N ({\rm cm}^{-2})}$. Detections are shown with $1\sigma$ error bars and upper limits are at $3\sigma$ confidence. For cases in which absorption column densities are measured in more than one line, the dispersion of individual measurements was used to determine the value listed.}

\tablenotetext{a}{This column density is the sum of the column densities for the O\,VI absorbers detected at $v_{\rm lsr} = -130$~\kms\ and $-70$~\kms\ (see Table~\ref{tab:pksparams} and Figure~\ref{fig:datafit}).}

\end{deluxetable}

\clearpage
\begin{deluxetable}{lccccccccc}

\tabletypesize{\scriptsize}
\tablecolumns{10}
\tablewidth{0pt}
\rotate

\tablecaption{Logarithmic Column Density Ratios\label{tab:hvc}}

\tablehead{\colhead{} & \colhead{$\bar{v}_{\rm lsr}$} & \colhead{} & \colhead{\underline{~\NHI~}} & \colhead{\underline{~\NSiII~}} & \colhead{\underline{~\NCII~}} & \colhead{\underline{\NCIV}} & \colhead{\underline{\NCIV}} & \colhead{\underline{\NOVI}} & \colhead{} \\ \colhead{Sight Line} & \colhead{(\kms)} & \colhead{\NHI\tablenotemark{a}} & \colhead{\NOVI} & \colhead{\NSiIV} & \colhead{\NCIV} & \colhead{\NSiIV} & \colhead{\NNV} & \colhead{\NCIV} & \colhead{$[Z/H]$\tablenotemark{b}}}

\startdata
PG~1259+593   & $-110$  & $19.92\pm0.01$          & $6.38\pm0.05$ & $\phm{-}2.03\pm0.22$ & \nodata              & $0.59\pm0.15$ & $> \phm{-}0.07$ & $\phm{-}0.31\pm0.09$ & $-1.00^{+0.19}_{-0.25}$\tablenotemark{c} \\
Mrk~279       & $-140$  & $19.49^{+0.06}_{-0.08}$ & $5.83\pm0.10$ & \nodata              & \nodata              & $0.67\pm0.20$ & $0.75\pm0.23$   & $-0.02\pm0.16$       & $-0.71^{+0.36}_{-0.25}$\tablenotemark{c} \\
PG~1116+215   & $+184$  & $17.82^{+0.12}_{-0.14}$ & $3.82\pm0.18$ & $\phm{-}0.76\pm0.07$ & $> \phm{-}0.81$      & $0.65\pm0.07$ & $> \phm{-}0.64$ & $\phm{-}0.26\pm0.05$ & $-0.66^{+0.39}_{-0.16}$ \\
HE~0226--4110 & $+175$  & $16.74\pm0.20$          & $3.40\pm0.22$ & $\phm{-}0.43\pm0.20$ & $> \phm{-}0.34$      & $> 0.84$      & $> \phm{-}0.23$ & $\phm{-}0.14\pm0.21$ & $-0.9\pm0.3$\tablenotemark{d} \\
PKS~2155--304 & $-140$  & $16.37^{+0.22}_{-0.14}$ & $2.57\pm0.22$ & $\phm{-}0.07\pm0.14$ & $\phm{-}0.36\pm0.11$ & $0.74\pm0.09$ & $> \phm{-}0.40$ & $\phm{-}0.32\pm0.05$ & $-0.47^{+0.15}_{-0.24}$ \\
HE~0226--4110 & $+193$  & $16.34\pm0.10$          & $3.05\pm0.15$ & $\phm{-}0.61\pm0.38$ & $-0.14\pm0.17$       & $0.46\pm0.32$ & $> -0.04$       & $-0.02\pm0.30$       & $-0.6\pm0.3$\tablenotemark{d} \\
HE~0226--4110 & $+~99$  & $16.29\pm0.05$          & $3.04\pm0.14$ & $> \phm{-}0.18$      & $> \phm{-}0.08$      & \nodata       & \nodata         & $> -0.10$            & $-0.6\pm0.2$\tablenotemark{d} \\
PG~0953+414   & $+125$  & $16.26\pm0.15$          & $2.52\pm0.17$ & $\phm{-}0.49\pm0.22$ & $< \phm{-}0.11$      & $0.65\pm0.31$ & $> -0.33$       & $\phm{-}0.75\pm0.22$ & $-0.8\pm0.2$\tablenotemark{d} \\
HE~0226--4110 & $+148$  & $16.21\pm0.10$          & $2.85\pm0.15$ & $\phm{-}0.54\pm0.40$ & $> \phm{-}0.29$      & $> 0.84$      & $> \phm{-}0.36$ & $\phm{-}0.00\pm0.24$ & $-0.4\pm0.2$\tablenotemark{d} \\
PG~0953+414   & $-150$  & $16.15\pm0.15$          & $> 2.66$      & $> \phm{-}0.60$      & $\phm{-}0.18\pm0.20$ & $< 0.58$      & \nodata         & \nodata              & $-0.6\pm0.2$\tablenotemark{d} \\
PKS~2155--304 & $-270$  & $15.23^{+0.38}_{-0.22}$ & $1.67\pm0.38$ & $< \phm{-}0.51$      & $< -0.56$            & $1.30\pm0.16$ & $> \phm{-}0.47$ & $\phm{-}0.00\pm0.07$ & $-1.20^{+0.28}_{-0.45}$ \\
{\bf \mrk}    & $+~94$  & $<14.85$                & $< 1.39$      & $\phm{-}0.23\pm0.08$ & $< -0.44$            & $0.95\pm0.19$ & $> -0.03$       & $\phm{-}0.42\pm0.61$ & \nodata \\
\enddata

\tablerefs{Column densities, ratios, and metallicities for sight lines other than \mrk\ were taken from the following sources: \citet{fox04,fox05} for Mrk~279; \citet{fox05} for HE~0226--4110 and PG~0953+414; \citet{collins03} for PG~1259+593; \citet{collins04} for PKS~2155--304; \citet{ganguly05} for PG~1116+215.}

\tablenotetext{a}{The logarithm of the H\,I column density, $\log{\NHI({\rm cm}^{-2})}$. Errors are quoted at $1\sigma$ confidence and upper limits are quoted at $3\sigma$ confidence.}
\tablenotetext{b}{The logarithm of the best-fit metallicity compared to Solar, i.e. $[Z/H] \equiv \log{(Z/H)_{\rm obs}} - \log{(Z/H)_{\Sun}}$.}
\tablenotetext{c}{The $[{\rm O\,I/H\,I}]$ value found by \citet{collins03}.}
\tablenotetext{d}{Metallicities from \citet{fox05} are quoted with 95\% confidence intervals; all other metallicities are quoted with $1\sigma$ error bars.}

\end{deluxetable}

%Figures
\clearpage

\clearpage

\clearpage

\end{document}